\documentclass[10pt, journal, compsoc, twocolumn]{IEEEtran}

\usepackage{multicol}

\usepackage{comment}
\usepackage{etoolbox}
 \usepackage{tabulary,booktabs}
 \usepackage[none]{hyphenat}
\usepackage[table,xcdraw]{xcolor}
\usepackage{multirow}
\usepackage{adjustbox}
\usepackage[table]{xcolor}
\usepackage{subcaption}
\usepackage{color}
\usepackage{xcolor}
\usepackage{hyperref}
\usepackage{tcolorbox}
\usepackage{tikz}
\usepackage[resetlabels,labeled]{multibib}
\newcites{S}{Literature Dataset} 

\input{lrsrefs}

\usepackage{microtype}
\usepackage[normalem]{ulem}
\AtBeginDocument{%
  \providecommand\BibTeX{{%
    \normalfont B\kern-0.5em{\scshape i\kern-0.25em b}\kern-0.8em\TeX}}}

\usepackage[resetlabels,labeled]{multibib}
\newcites{S}{Literature Dataset}

\ifCLASSOPTIONcompsoc
  \usepackage[nocompress]{cite}
\else
  \usepackage{cite}
\fi
\ifCLASSINFOpdf
\else
\fi

\begin{document}

%
\title{ A Systematic Mapping Study on Architectural Approaches to Software Performance Analysis}

\author{Yutong~Zhao, Lu~Xiao, Chenhao~Wei, Rick~Kazman, and~Ye~Yang
}

\newcommand{\SLR}{\textit{systematic literature review}}

\markboth{Journal of \LaTeX\ Class Files,~Vol.~14, No.~8, August~2015}%
{Zhao \MakeLowercase{\textit{et al.}}: A Systematic Mapping Review of Software Performance and Architecture Analysis Integration}

\IEEEtitleabstractindextext{%
\begin{abstract}

Software architecture is the foundation of a system's ability to achieve various quality attributes, including software performance. However, there lacks comprehensive and in-depth understanding of why and how software architecture and performance
analysis are integrated to guide related future research. To fill this gap, this paper presents a systematic mapping study of 109 papers that
integrate software architecture and performance analysis. We focused on five research questions that provide guidance for researchers
and practitioners to gain an in-depth understanding of this research area. These questions addressed: a systematic mapping of related
studies based on the high-level research purposes and specific focuses (RQ1), the software development activities these studies
intended to facilitate (RQ2), the typical study templates of different research purposes (RQ3), the available tools and instruments for
automating the analysis (RQ4), and the evaluation methodology employed in the studies (RQ5). Through these research questions, we
also identified critical research gaps and future directions, including: 1) the lack of available tools and benchmark datasets to support
replication, cross-validation and comparison of studies; 2) the need for architecture and performance analysis techniques that handle the
challenges in emerging software domains; 3) the lack of consideration of practical factors that impact the adoption of the architecture and
performance analysis approaches; and finally 4) the need for the adoption of modern ML/AI techniques to efficiently integrate architecture
and performance analysis.

\end{abstract}
}

\maketitle

\IEEEdisplaynontitleabstractindextext

%
\IEEEpeerreviewmaketitle

\section{Introduction}
\label{introduction}
Software architecture is defined as the set of software elements, their properties, and the relationships among them~\cite{bass2003software}. It is the foundation of a system's ability to achieve various quality attributes~\cite{franch1998putting}. Software performance is, for many systems, the most important quality attribute driving the design~\cite{zaman2012qualitative, nistor2013discovering}. Performance is the ability of a software system to perform its duties according to time constraints within its allowance of resources~\cite{zaman2012qualitative}. Poor performance can result in long execution times, unhappy users, and even system crashes~\cite{zaman2012qualitative, nistor2013discovering}. 
Performance problems may be introduced and incubate in the system in early software design and architecture decisions~\cite{becker2008coupled}. These decisions may lead to severe performance compromises in the system and may require significant effort to detect and fix later, often during software maintenance. Thus, architects should consciously address performance design considerations in their earliest software architecture design activities.

Due to the intrinsic connections between software architecture and performance, there is a substantial body of literature that integrates architecture and performance analysis~\cite{aleti2012software, olabiyisi2010survey, becker2006performance, balsamo2004model, arcelli2020exploiting, becker2012model, falessi2010applying}. The most thoroughly studied aspect is model-based performance prediction. That is, practitioners leverage software architecture models to predict the performance measures of a system in the early stages of the software development life cycle. A number of secondary studies have already summarized the state-of-the-art in this direction~\cite{olabiyisi2010survey, becker2006performance, balsamo2004model}. However, it remains unclear what are the other pillars of software architecture modeling and analysis that can facilitate software performance in practice. In particular, we are interested to study if there are methodologies that leverage architectural information to address software performance in the later stages of software development life-cycle, such as during software maintenance. 

The high-level objective of this study is to help researchers and practitioners gain a comprehensive and in-depth understanding of why and how software architecture and performance analysis are integrated in the current literature. Toward this objective, we searched and retrieved a total of 24901 potentially relevant studies, out of which 109 studies are retained in our final dataset. The process is guided by the well-established \textcolor{black}{systematic mapping study} process~\cite{kitchenham2011repeatability, kitchenham2015evidence}. We provide a landscape of the current state in the integration of software architecture and performance analysis. More specifically, provided answers to three research questions:

 \textit{RQ1: What are the different research purposes and facilitated development activities of integrating software architecture and performance analysis?} The 109 studies that integrate software architecture and performance analysis serve four objectives 1) model-based performance analysis (74 studies); 2) performance anti-pattern detection and resolution (16 studies); 3) profiling and comparison of architectural alternatives (11 studies); and 4) self-adaptive architecture for dynamic performance optimization (8 studies). We also provide a mapping of software development activities, including \textit{Implementation}, \textit{Deployment}, \textit{Operation}, and \textit{Maintenance}, with these four research purposes. The mapping provides guidance if one would like to identify the most relevant technique for their focused activity. It also highlights gaps that could be filled in future research.

\textit{RQ2: How are software architecture and software performance analysis integrated for different research purposes?} We summarized the typical study templates of the four fundamental purposes from the 109 studies, which offer a guide for reproducing existing studies and advancing the state-of-the-art.  Guidance regarding the key analysis components in the templates, including the architecture models, model annotation, performance models, and anti-pattern detection rules are provided.

\textit{RQ3: To what degree is the architecture performance analysis automated by available tools and instruments?} We provided a catalog of tools for architecture analysis and for performance profiling curated from the 109 studies. This can help researchers and practitioners identify resources for their own research, or for reproduction studies.


Finally, we discussed the limitations of existing research, from which a tentative future research roadmap with related research questions is shared for researchers and practitioners to refer to and build upon, including:
1) how to enhance the reproducibility and interoperability of existing research based on the techniques, tools, and datasets analyzed from existing studies? 2) How well do existing software architecture and performance analysis techniques apply to emerging domains, such as Web 3.0 and VR? What are the unique challenges with these new domains in the direction of architecture and performance analysis? 3) What are the most significant challenges and impacting factors on the practical usage of architecture and performance analysis techniques? And, 4) how to enable AI in architecture and performance analysis, in particular for mitigating the architecture analysis's reliance on experts and conquering the complexity and uncertainty of performance analysis?

In summary, the key contributions of this study include:
\begin{itemize}
  \item A comprehensive overview of the current state of research that integrates software architecture and performance analysis.
  
  \item A summary of available tools and instruments that support the integration of software architecture and performance research, including  tools for constructing and analyzing architecture models and performance models, as well as instruments for collecting performance metrics.
  
  \item A summary of limitations of the prior work, which motivates future directions that need more attention.
\end{itemize}

The rest of this paper is organized as follows: Section~\ref{sec:background} introduces the background  of software architecture and software performance;  Section~\ref{sec:approach} illustrates our methodology for preparing, retrieving, selecting, annotating, and synthesizing the relevant literature; Section~\ref{sec:data-overview} introduces the overall statistics of our dataset; Section~\ref{sec:results} presents the study results; Section~\ref{sec:implication} discusses the potential future research directions; Section~\ref{sec:related} discuss similar secondary studies; and Section~\ref{sec:conclusion} concludes. 
\section{background}
\label{sec:background}

In this section, we introduce background information about software architecture and software performance.

\subsection{Software Architecture}

\subsubsection{\textbf{Software Architecture Definition}} Bass \textit{et al.} define software architecture as \textit{``the structure or structures of the system, which comprise software elements, the externally visible properties of those elements, and the relationships among them.''}~\cite{bass2003software}. When describing software architecture, practitioners and researchers often treat software components as the composing elements~\cite{becker2008coupled}. 
A software component is a modular, portable, replaceable, and reusable set of well-defined functionality that encapsulates its implementation, and the component is exported as a high-level interface~\cite{becker2008coupled}. The externally visible properties of an architectural element refer to the assumptions that other elements can make of it, such as services it provide, its performance characteristics, shared resource usage, and so on~\cite{garlan2007software}. The architecture of a software system serves as a blueprint of the system, which involves the high level structure of software system abstraction~\cite{becker2008coupled}. The design of software architecture must be compatible with the functionality of the system, but is driven by its quality attribute requirements such as performance, reliability, scalability, availability, and flexibility, etc~\cite{becker2008coupled}.



\subsubsection{\textbf{Software Architecture Models}} 
Different stakeholders view software architecture differently, as discussed in the classic \textit{``4+1" view}~\cite{kruchten19954+}. To describe software architecture, practitioners use a variety of models and their representations, each serving distinct purposes. For instance, the Unified Modeling Language (UML) is an object-oriented language primarily used to visualize, specify, construct, and document the high-level structure and behavior of a software system~\cite{hofmeister1999describing, riva2004uml}. Architecture Description Languages (ADLs) are another class of models that provide precise syntactic and semantic descriptions for software architecture~\cite{smith2003software}. They support the formal specification of software features such as processes, threads, data, and sub-programs, as well as hardware components like processors, devices, buses, and memory. Other software architecture models include the Palladio Component Model, Message Sequence Charts, and the Descartes Modeling Language~\citeS{li2016evaluating, du2015evolutionary, arcelli2015control, wu2015exploring}. The Palladio Component Model aims to predict performance, reliability, and maintainability of software architectures in early development stages, while Message Sequence Charts display interactions between processes or objects in real-time systems, and the Descartes Modeling Language is used for performance prediction and system quality evaluation for self-adaptive software systems.

\subsection{Software Performance}

\subsubsection{\textbf{Software Performance Engineering}}

According to Smith \textit{et al.}, performance concerns how well a software system or its components meets the requirements for timeliness~\cite{smith2002performance}. Timeliness encompasses two important dimensions: response time and throughput~\cite{haring2003performance, balbo1979approximate}. Response time is the time required to respond to events, while throughput is the number of events processed during a time interval~\cite{balbo1979approximate}. Zaman \textit{et al.} further extends the performance concerns by considering resource utilization as another aspect of software performance~\cite{zaman2012qualitative, zaman2011security}. The resources of interest usually include: 1) hardware resources, such as CPU, disk I/O, and memory; 2) logical resources, such as buffers, locks, and semaphores; and 3) processing resources, such as threads and processes~\cite{zaman2012qualitative, nistor2013discovering, cortellessa2005far, woodside2007future}. Moreover, recent studies also address other system characteristics, such as the data transmission loss ratio~\citeS{sunardi2019mvc, trubiani2018exploiting}, energy consumption~\citeS{camara2020quantitative}, and network latency~\citeS{smith2017spe}.

To ensure these performance concerns are met throughout the development process, practitioners must incorporate performance considerations from the outset, rather than deferring them until testing—a practice referred to as the "fix-it-later" approach. To alleviate the problems caused by this delayed approach, Smith \textit{et al.} proposed \textit{Software Performance Engineering}\cite{smith2003software}, a systematic, quantitative approach initiated in the early stages of the software development life cycle, and continuing through architecture design, implementation, and maintenance\cite{woodside2007future}. This performance engineering life cycle mirrors the software development life cycle~\cite{smith2007introduction}. It emphasizes the analysis of requirements regarding performance, scalability, stability, and reliability during requirement gathering and analysis~\cite{woodside2007future}. During the design phase, it deploys modelling techniques like \textit{Queuing Networks}, \textit{Place/Transition (Petri) Nets}, and \textit{Stochastic Process Algebra}\cite{balsamo2004model} to validate design assumptions concerning performance. When it comes to implementation, this approach provides guidelines for validating and reporting on the performance of the code\cite{woodside2007future}. Lastly, in the maintenance phase, it focuses on monitoring the performance metrics of a system, thoroughly evaluating their compliance with the performance requirements, and proposing potential optimization recommendations~\cite{brosig2014architecture, smith2002performance}.



\subsubsection{\textbf{Software Performance Testing and Profiling}} Performance testing is one of the most thoroughly studied approaches to addressing performance concerns. The purpose of performance testing is to identify bottlenecks in software systems~\cite{weyuker2000experience}.  Performance testing executes a system and constructs a profile of it, in terms of responsiveness and stability under various workloads~\cite{smith2002performance}. Avritzer and Weyuker made notable contributions to this area with their work on the automatic generation of load test suites and the assessment of the resulting software~\cite{avritzer1995automatic}. There are four major types of performance testing methodologies: load testing, stress testing, endurance testing, and spike testing~\cite{nistor2013discovering, bass2003software}. Load testing evaluates the behavior of a software system under specific workloads. Stress testing executes and profiles the system under extreme workloads to discover the maximum capacity of the system. Endurance testing executes and profiles the system under continuous workload. The purpose is to determine whether the system can scale up to support enduring and increasing workloads. Spike testing determines whether a system can sustain a sudden increase in workload~\cite{woodside2007future}.

During performance testing, practitioners leverage profiling tools to keep track of metrics such as response time, throughput, and resource utilization~\cite{patel2012software}. \textcolor{black}{Avritzer and Weyuker also investigated metrics for architectural assessment in performance testing, enhancing the understanding of the metrics necessary for assessing software architecture~\cite{avritzer1998investigating}.} These tools are available for many platforms, programming languages, and execution environments, with different advantages~\cite{mustafa2009classification,noauthor_7_2020}. For example, \textit{WebLoad} can generate real-life and reliable workload scenarios for testing complex systems~\cite{sobel2008cloudstone}. \textit{LoadNinja} has the highest coverage for performance testing~\cite{srivastava2021software}. \textit{LoadView} can be applied to real-life browsers and web applications~\cite{san2017water}. \textit{StresStimulus} can detect hidden concurrency errors by measuring performance metrics, such as network latency and data transmission loss ratio~\cite{markey2013performance}. \textit{Apache JMeter}~\cite{halili2008apache} is the most widely-used performance testing and profiling tool for \textit{Java} projects and it has been integrated to many IDEs. \textit{SmartMeter} can automatically generate a performance assessment report~\cite{prasad2013smart}. \textit{Rational Performance Tester} is a powerful performance testing tool developed by \textit{IBM}~\cite{applyingusing}, which supports load testing that involves multiple users and generates a comprehensive performance assessment report.  

\section{Study Process}
\label{sec:approach}
This study followed the guidelines for \SLR{} studies in  software engineering~\cite{kitchenham2011repeatability, kitchenham2015evidence}. We followed the process suggested by Kitchenham et al.~\cite{kitchenham2015evidence} with study planning (i.e. defining the RQs), literature searching, study selection, extracting study data, and data aggregation and synthesis. In addition, we followed the guidelines in~\cite{kitchenham2011repeatability} for ensuring stability and reliability. We recorded detailed data from each step, including  initial search results,  notes for the selection process, and the data annotation and synthesis, to maintain a clear chain of evidence for our findings. The detailed data is available on a public repository: \url{https://sites.google.com/view/arch-perf-data-repo/}

\begin{figure}[h]
  \centering
  \includegraphics[width=\columnwidth]{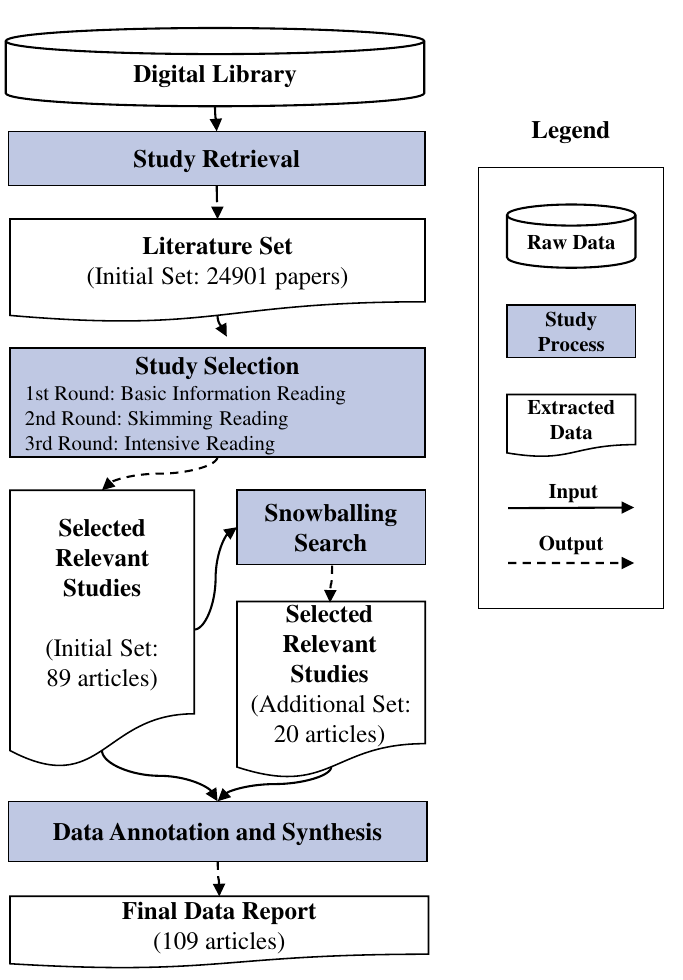}  
  \caption{Overview of Study Process}
  \label{fig:approach}
\end{figure}

Figure~\ref{fig:approach} illustrates the five steps in this study:
\begin{enumerate}
    \item \textit{\textbf{Study Preparation:}} This step defined the research questions.
    \item \textit{\textbf{Study Retrieval:}} We defined the search strings and retrieved the initial set of papers from major digital libraries.
    \item \textit{\textbf{Study Selection}:} We conducted three rounds of selections to identify the most relevant literature. 
    \item \textit{\textbf{Snowballing Search:}} We further examined the references of the literature retrieved from Step 2 to identify additional studies.
    \item \textit{\textbf{Data Annotation and Synthesis:}} We read each paper in the final set, annotated information on each paper that was relevant to our research questions, and synthesized the results to answer the research questions.
\end{enumerate}

\subsection{Study Preparation: Defining Research Questions}
\label{sec:rqs}
The first step was defining the research questions, which guided the rest of the study.
    
\textit{\textbf{RQ1:} \textbf{What are the different research purposes facilitated development activities of integrating software architecture and performance analysis?}} Model-based performance prediction, leverages software architecture models for predicting performance metrics of a system, is the most well-known motivation for integrating architecture and performance analysis. However, it remains unclear whether exist and what are the other important themes that motivate the integration of architecture and performance analysis.In addition, this RQ investigate the development activities that related studies target at and facilitate. A prior study\textcolor{black}{~\cite{balsamo2004model}} reported that the model-based prediction techniques were mostly applied during the software architecture design. This provides early performance assessment and enables optimization of the system architecture design before the implementation starts. However, it was not clear whether the different research purposes usually maps to different development activities, especially activities other than software architecture design reported in the prior study.
The motivation of this RQ is to provide practitioners a comprehensive categorization and mapping of existing studies based on their objectives and the specific development activities that they concern. As such, this mapping can guide future researchers and practitioners to the studies that best fit their interests.


\textit{\textbf{RQ2:} \textbf{How are software architecture and software performance analysis integrated for different research purposes?}} This RQ aims to address ``how'' software architecture and performance analysis are integrated for different objectives. We aim to provide reference research templates of the different research purposes identified in RQ1. Each template should summarize the common study steps and their workflow from papers of the same purpose. 

\textit{\textbf{RQ3: To what degree is the architecture performance analysis automated by available tools and instruments?}} This RQ aims to summarize tools and instruments that are used to automate the architecture and performance analysis. In particular, we aim to reveal what is the extent of manual effort reported in the literature. For summarizing the tools, we present two parts: 1) RQ-3.1: What are the available architecture and performance modeling and analysis tools? and 2) RQ-3.2: What are the performance profiling instruments for collecting dynamic performance metrics, and which metrics are collected? This RQ helps practitioners to quickly identify relevant tools or techniques for their purposes.

\subsection{Study Retrieval}
Our search string was defined as follows:
\begin{tcolorbox}[center,size=title,width=0.95\linewidth, opacityfill=0.15]
\textit{(software) \textbf{AND} (architecture \textbf{OR} architectural \textbf{OR} architecting \textbf{OR} design \textbf{OR} structure) \textbf{AND} (performance)} 
\label{}
\end{tcolorbox}
The first keyword, \textit{``software''}, establishes the overall focus on the field of software engineering. Originally, we included terms such as \textit{``architecture''}, \textit{``architectural''}, \textit{``architecting''}, \textit{``design''}, \textit{``structure''}, and \textit{``behavior''} to encompass various aspects of software architecture. 

The first keyword, \textit{``software''}, establishes the overall focus on the field of software engineering. Originally, we included terms such as \textit{``architecture''}, \textit{``architectural''}, \textit{``architecting''}, \textit{``design''}, \textit{``structure''}, and \textit{``behavior''} to encompass various aspects of software architecture. However, introducing ``\textit{behavior}'' retrieved an overabundance of unrelated papers primarily covering human behavior studies, hence, it was excluded to maintain specificity. Further exploration of architecture-related terms, including ``\textit{component-based system}'', ``\textit{distributed system}'', and ``\textit{service-oriented system}'', did increase the total number of papers, but our analysis revealed that all these papers were already included within the scope of the terms ``\textit{architecture}'', ``\textit{design}'', or ``\textit{structure}''. Therefore, these additional terms were deemed redundant and excluded. This streamlined search string demonstrated its effectiveness and efficiency in our subsequent experiment, where it successfully retrieved relevant papers without introducing unnecessary noise or omitting key studies.However, the term ``\textit{behavior}'' could also represent the broader human behavior related topics and led to large noise. We removed ``\textit{behavior}'' out of the final search string.

We conducted the search on seven digital libraries: 1) \textit{IEEEXplore}, 2) \textit{ACM Digital Library}, 3) \textit{Web of Science}, 4) \textit{Springer Digital Library}, 5) \textit{Elsevier ScienceDirect}, 6) \textit{Wiley Online Library}, and 7) \textit{Google Scholar}. For each library, we searched the document title, keywords, and abstract, to comprehensively retrieve potential relevant studies. The search range was January 2010 to December 2020. The guideline in \cite{kitchenham2015evidence} specifies the first six libraries for software engineering literature review. We included \textit{Google Scholar} following the practice in~\cite{aleti2012software, olabiyisi2010survey}, to be comprehensive. 
To the best of our knowledge, these seven libraries should be comprehensive in covering literature from major software engineering venues.

\subsection{Study Selection}
Next, we conducted three rounds of selection to determine papers in our final study dataset. Table~\ref{tbl:criteria} lists the inclusion and exclusion (I\&E) criteria.  
\begin{table}[h]
\normalsize
\centering
\caption{Inclusion and Exclusion Criteria}
\begin{adjustbox}{width=\columnwidth}
\begin{tabular}{||c l||} 
 \hline
ID & Inclusion Criteria \\ [0.5ex] 
 \hline
 I1 & The paper is peer-reviewed. \\
 I2 & The paper is written in English. \\
 I3 & The paper contains six or more pages. \\
 I4 & The paper is published in an international conference, journal or symposium. \\
  \hline\hline
 ID & Exclusion Criteria \\ [0.5ex] 
 \hline
 E1 & A previous version of the paper whose extended version has been included.  \\
 E2 & The paper is a secondary study (literature review) of existing techniques/approaches. \\ 
 E3 & The paper focus on performance but does not consider software architecture.\\
 E4 & The paper relates to architecture but does not focus on software performance.  \\
 E5 & The paper focuses on hardware-related performance. \\
 E6 & The paper is not a full paper, e.g. missing evaluation. \\
 [1ex] 
 \hline
\end{tabular}
\end{adjustbox}
\label{tbl:criteria}
\end{table}

\subsubsection{\textbf{1st Round: Basic Information Reading}} This round examined the basic information---the publication venue, the number of pages, and the title of each paper. 

We first applied a set of objective inclusion criteria, including I1, I2, I3, and I4 (defined in ~Table \ref{tbl:criteria}) to retain papers that met certain quality specifications.  I1 ensured that only peer-reviewed papers were included. For example, white papers, technical reports, and thesis/dissertations were not included. I2 ensured that only papers written in English were included. I3 ensured that papers with six or more pages were included. The motivation was to only include full research papers with elaborated research methodology and evaluation. Although most premier software engineering venues, such as ICSE, FSE, and ASE, have a ten page requirement for full research papers, we loosen this to  six or more pages to make sure that relevant papers from venues with a different requirement were also included. Based on our observations, papers less than six pages were not likely to contain sufficient detail. I4 ensured that we only included papers published in an international conference, journal or symposium. Studies published at regional venues, such as ``\textit{4th India Software Engineering Conference}''~\cite{kumar2011architectural} and ``\textit{Journal of Zhejiang University}''~\cite{bushehrian2010automatic} were removed by this criterion. 

Next, we scanned the title of each paper, and applied two exclusion criteria, E2 and E5. E2 excluded papers that were secondary studies, such as literature reviews or surveys. For example, we excluded the paper \cite{mahdavi2017systematic} ``\textit{A systematic literature review on methods that handle multiple quality attributes in architecture-based self-adaptive systems}''. E5 excluded papers that focused on hardware performance such as multi-core processors. For example, ``\textit{Performance Evaluation of a Hybrid Computer Cluster Built on IBM POWER8 Microprocessors}''~\cite{mal2019performance} apparently focused on hardware. Note that it was not always possible to apply E2 or E5 simply based on the title of the paper. As discussed later, these two criteria were also used in the next two rounds with greater understanding of a paper. 

\subsubsection{\textbf{2nd Round: Skimming}} In this round, we skimmed through the abstract, introduction, and conclusion of each remaining paper, which helped us to grasp the key theme of the study. We applied the exclusion criteria, E1, E3, E4, and E5 based on the skim reading. E1 excluded papers which had an extended version in our search results. As an example, \citeS{happe2014stateful} indicated that ``\textit{The paper is an extension of our previous work~\cite{kapova2010state}}'', and thus ~\cite{kapova2010state} was removed. 
Criteria E3 and E4 excluded papers that focused either on architecture or on performance, but not the integration of the two. For example, the study \cite{eichelberger2015adaptive} was excluded by E3 because it only profiled the performance of big data applications. The study did not investigate the software architecture of the system. As an example for E4, the study \cite{guerriero2016towards} was excluded since it investigated the effort for the architecture design of software systems, while the performance of the systems was not a key focus. Criterion E5 removed papers that focused on hardware performance. For example, we realized that ``\textit{Modeling performances of concurrent big data applications}'' \cite{castiglione2015modeling} analyzed the speed and power consumption of multi-core processors in data center infrastructures.

\subsubsection{\textbf{3rd Round: Intensive Reading}} 
In the final round, we carefully reviewed each remaining paper. Here, we applied all the exclusion criteria (E1 to E6). Note that some of the criteria (\textcolor{black}{E1}, E2, E3, E4, and E5) were already applied in previous rounds based on partial reading. However, in the previous two rounds we always preferred inclusion over exclusion for cases lacking strong confidence. The goal was to make sure that we did not exclude papers recklessly. Thus, these five criteria were evaluated again based on the updated understanding of the full content in the paper. 
For E1, some papers (e.g., \citeS{arcelli2015performance}, \citeS{happe2014stateful}) clearly stated that they were extensions to previous versions in the introduction, and thus E was applied in skim reading. While, some papers, e.g. \citeS{brunnert2017continuous}, clarified this in the background or even study approach sections, and thus were excluded by E1 in intensive reading.
In comparison, E6 was evaluated for the first time, which excluded papers that did not contain any sort of evaluation or case study. As examples, \cite{spinner2016reference, pagliari2019extent} were excluded due to E6, since they only presented conceptual solutions without any evaluations or experiments.

\subsection{Snowballing Search}
As shown in Figure~\ref{fig:approach}, the papers selected after the three rounds were called the \textit{Initial Set}. Next we performed backward snowballing based on papers in this \textit{Initial Set}. Our goal was to retrieve additional  papers that were missed in the search process. For example, there may exist  papers that were not included in the seven digital libraries used in our study. As discussed later, we were able to identify an \textit{Additional Set} with 699 papers by tracing the references of each paper in the \textit{Initial Set}. Of  note, this was after excluding papers published before January 2010 and papers that were already in our initial search. We repeated the three rounds of selection as described above on the \textit{Additional Set} from snowballing. This ensured that all the retrieved papers, through initial search or snowballing, were evaluated following the same process and criteria. Therefore, it is reasonable for us to claim that our study captures related studies comprehensively. 




\subsection{Data Annotation and Synthesis} 
\begin{table*}[t]
\normalsize
\caption{Annotated Data Items}
\begin{adjustbox}{width=\textwidth,center}
\centering
\begin{tabular}{|c l l c|} 
 \hline
ID & Data Item & Description  & Correspond to \\ [0.5ex] 
 \hline\hline
 D1 & Document Title & The title of the article. & \multirow{5}{*}{General}\\
 D2 & Publication Venue & Published in which Conference, journal, or workshop, etc. &  \\
 D3 & Authors & The author names. & \\ 
 D4 & Publication Year & The year when the study is published. &  \\
 D5 & \# Pages & The number of pages in the paper. &  \\ \hline \hline
 D6 &  Purpose Category & Mapping of this paper to the hierarchical categorization of all papers. & \multirow{4}{*}{RQ1} \\
 D7 &  Motivation & The extracted description of the motivation of the paper. & \\ 
 D8 & Purpose Code & The codes for key study purposes used by two authors independently by open coding. &  \\ 
 D9 & Development Activity & The software development activities the proposed techniques aim to facilitate &  \\ \hline\hline
 D10 & Study Process  & The summary of key study steps & \multirow{6}{*}{RQ2} \\
 D11 & Architecture Model & The architecture model employed in the study. &  \\
 D12 & Performance Model & The performance model employed in the study. &  \\
 D13 & Annotated Parameters & The performance parameters being annotated to architecture models &  \\
 D14 & Performance Metrics & The metrics used to measure software performance in a study. &  \\ 
 D15 & M2M Integration & The integration of architecture and performance model analysis. &  \\ \hline\hline
  D16 & Modelling Tools & Tools that are used for creating and analyzing architecture and performance models & \multirow{3}{*}{RQ3} \\
 D17 & Profiling Tool & The profiling tool for collecting the performance metrics. &  \\
 D18 & Manual Effort & The manual effort involved in the research steps. &  \\ \hline\hline
 D19 & Future Work & The discussed limitations and future work directions. & \textcolor{black}{Future Directions} \\ 
 [1ex] 
 \hline
\end{tabular}
\end{adjustbox}
\label{tbl:annotation}
\end{table*}
In this step, we carefully read each paper and annotated information that correspond to the RQs discussed in Section~\ref{sec:rqs}. Table~\ref{tbl:annotation} lists the detailed data items we annotated. There are two types of data items: 1) general information, such as the title, publication type, authors, publication years and the number of (\#) pages---listed as D1 to D5 in the first half of the paper; and 2) specific information that maps to the RQs --- listed as D6 to D18.

\begin{table*}
\caption{Examples of Manually Open Coding for Study Purpose Categorization}
\begin{adjustbox}{width=\textwidth,center}
\centering
\begin{tabulary}{\textwidth}{@{}*3{|L}*2{|L|}*2{|L}}
 \hline
 \multicolumn{1}{|>{\large}c|}{Reference} & \multicolumn{1}{|>{\large}c|}{Extracted Motivation (D7)} & \multicolumn{1}{|>{\large}c|}{Author-1 Coding (D8)} & \multicolumn{1}{|>{\large}c|}{Author-2 Coding (D8)} \\ [0.5ex] 
 \hline\hline
 \citeS{etxeberria2014performance}  & \textcolor{black}{Uncertainty is particularly critical in the performance domain when it relates to values of parameters such as workload, operational profile, resource demand of services, service time of hardware devices, etc. The goal of this paper is to explicitly consider uncertainty in the performance modelling and analysis process. }
 & Recognizing and managing the uncertainties in software models & Model-based performance evaluation to handle uncertainty. \\ \hline
 \citeS{brosig2014architecture} & Current approaches to modeling the context of software components are not suitable for use at run-time. In this paper, we aim to propose a performance meta-model (DMM) for run-time performance prediction, specifically designed for use in online scenarios. & (1) Model-based prediction; (2) Run-time predictions. & Model-based run-time prediction using DMM. \\ \hline
\end{tabulary}
\end{adjustbox}
\label{tbl:open-coding}
\end{table*}

For RQ1, we first extracted text describing the motivation of each paper as D7. Next, two authors independently performed open coding, D8, which summarized the key objective of the study based on D7. Finally, we merged the codes from D8 to derive a hierarchical categorization of the papers, and mapped each paper to a category recorded as D6. Table~\ref{tbl:open-coding} shows examples of the annotated motivation (D7) and the codes (D8) generated by the two authors. 


After the annotation and open coding were done, the two authors constructed a hierarchical categorization of the papers by merging their codes. They first came up with the most high-level categorization of the papers based on the most fundamental objectives. For example, the codes of \cite{brosig2014architecture} suggested the key theme of predicting and evaluating performance based on architecture/performance modeling; thus they were merged to derive the category called \textit{Model-based Performance Prediction}. Similarly, \citeS{trubiani2014exploring} derived another category called \textit{Performance Anti-pattern Detection and Solution}. Next, the two authors further derived sub-categories based on the specific focuses (if available) of each paper. For example, both \citeS{eismann2018modeling} and \citeS{brosig2014architecture} specifically focused on \textit{``run-time''} prediction reflected in the codes generated by the two authors. Thus they led to a sub-category, named \textit{Run-time Prediction}.


In the process of deriving the sub-categories, two authors had disagreements on nine papers. All authors were invited to read these papers in full, and each author summarized the key theme of the paper independently. Then, the team employed the Delphi technique in a team discussion to reach consensuses on these nine papers. 

In addition, we annotated information (D9) about the development activities that the proposed approach was intended to facilitate. For example, paper~\citeS{gribaudo2018performance} indicated that ``\textit{we proposed a modelling approach that provides fast evaluation to support design choices}''. Thus, this paper~\citeS{gribaudo2018performance} aimed at facilitating the software design. Some papers contributed techniques that may facilitate multiple development activities. 

For RQ2, we summarized the study process (D10) of each paper. This was usually based on the description of the proposed technique in the approach section of the paper. For example, for paper~\citeS{gomez2014performance}, the study process was annotated as ``\textit{1) Model Acquisition: modeling the system architecture using UML diagrams; 2) Model Annotation: use UML-MARTE to annotate the properties such as workload and resource demand; 3) Transformation: use ArgoSPE  to translate UML into GSPN; 4) Performance Analysis: calculate response time, scalability, and resource utilization by perf model simulation.}''. In addition, we also recorded the detailed information in the study process, including  the employed architecture model (D11), the performance model (D12), performance parameters in the modeling (D13), performance metrics being analyzed (D14), and the technique (D15) to integrate the analysis of architecture and performance models.

The study templates of different research purposes were synthesized by identifying and aggregating common research steps in the process annotations. For example, model-acquisition was an obvious common step for all the studies that leverage architecture modeling for performance prediction. Some papers retrieved architecture models based on the design documents~\citeS{gribaudo2018performance, arcelli2015control, gomez2014performance, wu2015exploring, werle2020data, requeno2017performance, kross2017model, nalepa2015model, pinciroli2021model, smith2017spe, nguyen2017parad, maddodi2020aggregate, yasaweerasinghelage2018predicting, noorshams2014enriching, walter2017expandable, brosig2014quantitative, woodside2014transformation, eshraghian2015performance, li2016evaluating}, while other papers recovered software architecture models from the existing systems by using reverse engineering techniques~\citeS{etxeberria2014performance, brosig2014architecture, trubiani2015exploiting, eismann2018modeling}. Of particular note, papers may use different terms for the same process. 

Furthermore, each paper may contain more fine-grained sub-steps. As an example, \citeS{noorshams2014enriching} presented a classification of performance-related parameters for annotating architecture models of a system. This sub-step categorized the performance-influencing factors as workload-specific characteristics (e.g., number and time interval of requests) and system-specific characteristics (e.g., resource demand of a specific system component), for the higher-level purpose of architecture model annotation. We considered this classification as a part of the model annotation step in the high-level study template. In addition, some papers may contain unique steps that did not recur in other studies. We did not include such unique steps as a part of the high-level research template. 


RQ3 focused on collecting the information of tools that automate  architecture and performance analysis. We recorded any tools that were used or developed for the analysis (D16). We also recorded any profiling tools (D17) that were used to collect the software performance metrics at run-time. We noted any explicitly mentioned manual effort (D18) in each study. 


Finally, to explore future research directions, as discussed in Section~\ref{sec:future-directions}, we annotated the limitations and future directions (D19) discussed in the papers.

\subsection{Threats to Validity}

In this section, we discuss the threats to validity of this study, following the guidelines provided in~\cite{zhou2016map}.

\subsubsection{Threats to Construct Validity}
\label{sec:construct-threat}

First, we cannot guarantee that the search string as constructed captured 100\% of all possible related studies --- especially those that used less common terms for software architecture or performance. This may impact the comprehensiveness of the results in this paper. To mitigate this problem, we included synonyms, alternatives and hypernyms in the search string to represent the concept of architecture. For instance, the keywords that are related to architecture include ``\textit{architectural}'', ``\textit{architecting}'', ``\textit{design}'', and ``\textit{structure}''. Plus, we also applied an additional snowballing search step to include the potentially missed studies in the first round of search. Thus, it is reasonable to assume that our study has covered most of the related literature in this field. 

In addition, we acknowledge that our study process did not include the formal quality assessment step suggested in ~\cite{kitchenham2015evidence}. This step was to ensure that the literature review was based on high quality studies. This required the definition of a formal assessment checklist and a scoring schema based on the checklist , to be performed on each individual study. However, we would like to argue that some of the inclusion and exclusion criteria, including I1, I3, I4, and E6 in Table~\ref{tbl:criteria}, served as filters to ensure the selected papers met reasonable quality standards. In particular, E6 filtered out preliminary studies whose validity was not evaluated. We did not define any further assessment criteria, such as the soundness of the proposed approaches and experiment results. The reason was that the retrieved literature was rich in architecture models, performance models, model transformation, and performance profiling tools/techniques. An in-depth assessment of  study quality was beyond the scope of this literature review.


\subsubsection{Threats to Conclusion Validity}
Although we followed the guidelines in~\cite{kitchenham2011repeatability} for ensuring  stability and reliability requirements, we cannot guarantee that our study results were free of any potential personal biases. The study results were potentially impacted by our  biases in conducting the selection, data annotation, and data aggregation and synthesis. This is a threat to conclusion validity.  However, we believe that following our replication package~\url{https://sites.google.com/view/arch-perf-data-repo/}, another researcher would be able to replicate our study following the same process and would reach consistent results and conclusions.  

To best mitigate personal bias in the selection process we employed two researchers to conduct the selection in \textit{Skimming Reading} and \textit{Intensive Reading} following the inclusion and exclusion criteria in Table~\ref{tbl:criteria}. The two researchers agreed on most papers. The \textit{Inter-Rater Reliability} of the final decision after these two rounds of selection was 86\%. Next, to  avoid inaccurate or incomplete data annotation, the data items in Table~\ref{tbl:annotation} were provided with detailed descriptions in our data repository. The data annotation was initially completed by the first author, and then checked by the second and third authors to high-light disagreements. The three authors held group discussions through online chat and real-time meetings to resolve disagreements. Any remaining disagreement after the review of the three authors was managed through group discussion. Furthermore, to best mitigate potential biases in the data synthesis, two researchers worked on analyzing the raw annotation data, and presented and discussed the derived results with all authors. In particular, the categorization of research purposes in RQ1 emerged from the data following an open coding approach, conducted by two authors independently, and then  extensively discussed and iteratively refined by the entire research team.

\subsubsection{Threats to Internal Validity}
The validity of our study results in RQ3 (research templates), RQ4 (available tools and instruments), and RQ5 (evaluation) are subject to a threat to internal validity  associated with the selected papers. The results for RQ3, RQ4, and RQ5 were synthesized from data annotation of the information items from the literature. If the authors of the original study failed to accurately provide the details of their study, this would impact our study results. For example, if the authors of a paper did not describe certain steps in their study which were not the key study contribution, this may impact our aggregation of the typical research templates in RQ3. Similarly, if the authors did not report the tools or instruments they used for architecture modeling and performance analysis, our results in RQ4 would be impacted. This threat was associated  with the retrieved literature, and was largely out of our control, but is presumably mitigated by the large number of papers studied.

\subsubsection{Threats to External Validity}
This study potentially suffers from threats to external validity. First, we focused research published between 2010 and 2020. We cannot guarantee that the findings derived from this range could be generalized to studies published before 2010 or after 2020. In addition, we cannot guarantee that the tools and techniques identified for software architecture modeling and analysis were representative of the tools and techniques used in a more general setting without performance analysis. Likewise, we summarized instruments for performance profiling in RQ4, but we cannot guarantee that these findings were representative of all the performance profiling tools that were used independent of architecture analysis.






\color{black}

\section{Study Results}
\label{sec:results}

\subsection{Overview of Study Dataset} \label{sec:data-overview}

After searching the seven digital libraries using the search terms, we retrieved a pool of 24901 papers in total. In the 1st round, i.e., basic information reading, we retained 1039 articles. In the second round, i.e., skimming, we retained 313 articles. In the third round, intensive reading, we retained 89 articles as our \textit{Initial Set}. Next, we applied a snowballing search process starting from these 89 articles. We collected a pool of 793 additional articles, after excluding papers before 2010 and removing duplicates. After repeating the three selection rounds on the these articles, we identified an additional 20 articles--- 195 articles retained after the first round; 111 articles retained after the second round; and 20 articles after the third round. Thus, there were a total of 109 (89+20) papers in the final set.

The majority, 68\%, of the studies were peer-reviewed full conference papers. 22\% of the studies were journal articles. 10\% of the studies were peer-reviewed workshop papers. Table~\ref{tbl:publication} lists the details of the publication venues. Those contributing only one study were combined as ``\textit{Others}". The venues that had the largest number of related papers were either specialized in software architecture, such as \textit{QoSA}~\footnote{QoSA is merged into ICSA after 2016.}, \textit{ICSA}, and \textit{ECSA} or specialized in software performance, such as \textit{ICPE}.

\begin{table}[h]
\normalsize
\caption{Publication Venues}
\begin{adjustbox}{width=\columnwidth}
\begin{tabular}{||l c c c c ||} 
 \hline
Publication Venue & Abbreviation & Type & No. & \% \\ [0.5ex] 
 \hline\hline
Inter. Conf. on Performance Engineering & ICPE & Conference & 13 & 12\% \\
Inter. Conf. on Quality of Software Architectures & QoSA & Conference & 12 & 11\% \\
Inter. Conf. on Software Architecture & ICSA & Conference & 10 & 10\% \\ 
European Conf. on Software Architecture & ECSA & Conference & 7 & 7\% \\
Inter. Conf. on Software Engineering & ICSE & Conference & 3 & 3\% \\ \hline

Journal of Systems and Software & JSS & Journal & 4 & 4\% \\
IEEE Transaction on Software Engineering & TSE & Journal & 3 & 3\% \\

Inter. Workshop on Software and Performance & WOSP & Workshop & 2 & 2\% \\ \hline \hline

\textit{Others} &  & & 43 & 41\% \\ \hline 
\end{tabular}
\end{adjustbox}
\label{tbl:publication}
\end{table}

\subsection{RQ-1: Study Purpose Categorization}\label{sec:rq1}

\begin{figure*}[t]
    \centering
    \includegraphics[width=1.8\columnwidth]{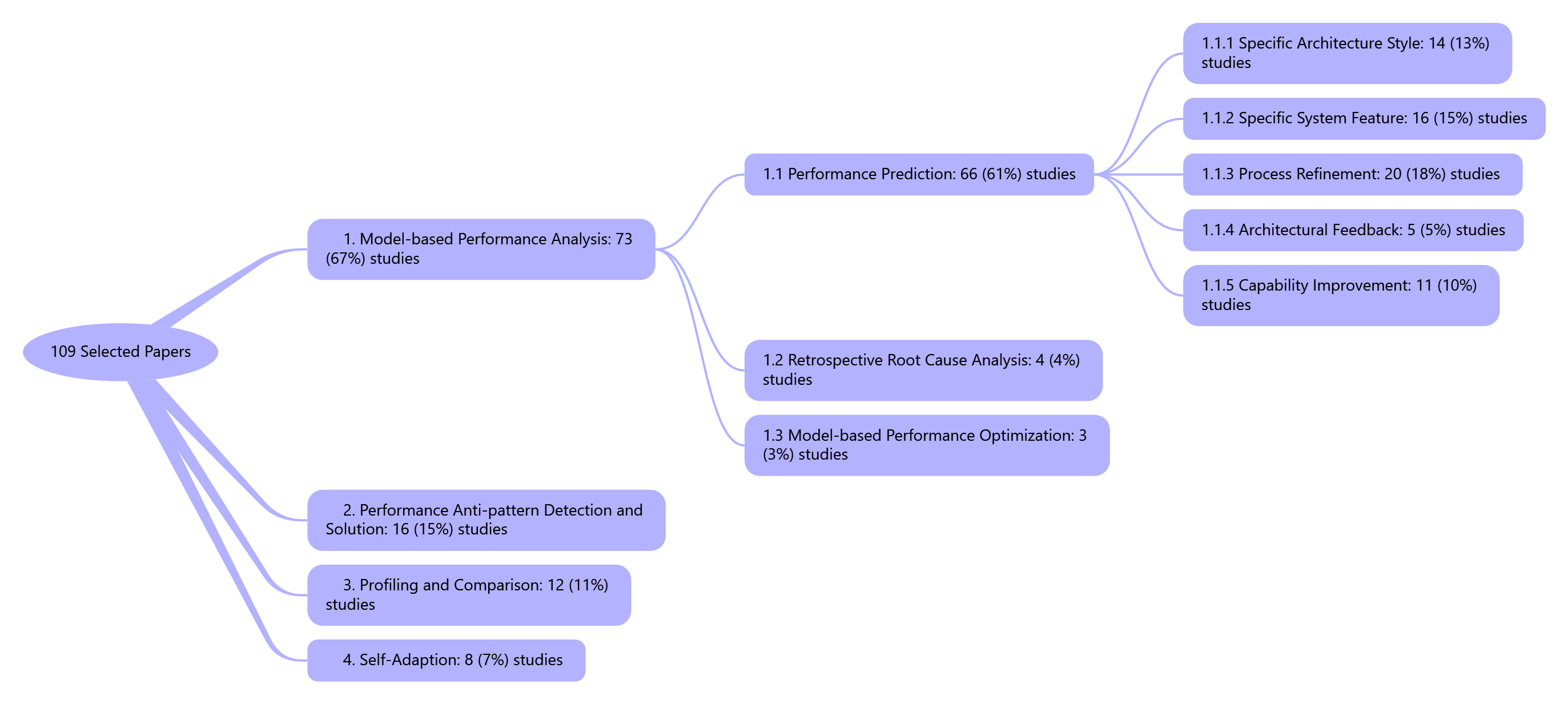}
    \caption{Study Categorization based on Purposes (RQ1) }
    \label{fig:purpose}
\end{figure*}

Figure~\ref{fig:purpose} illustrates our categorization of the 109 studies based on objectives and specific focuses. Overall a majority, 73 (67\%), of the studies leveraged architecture modeling for performance analysis, including performance prediction, root cause analysis, or architectural optimization. Given the large number of studies in this category, they were further sub-categorized based on the focus of each study. The other 36 (34\%) studies were categorized into three purposes: performance anti-pattern detection and resolution, performance profiling and comparison of alternative architectural solutions, and constructing software systems with self-adaptive architectures. 

Next, we discuss each category in detail.

\subsubsection{\textbf{Category 1: Model-based Performance Analysis}} 
This is the largest category, with 73 (67\%) studies. These studies were sub-categorized into: 1.1 Model-based Performance prediction; 1.2 Retrospective Root Cause Analysis of performance issues; and 1.3 Model-based Architectural Optimization.

\paragraph*{\textbf{Category 1.1 Model-based Performance Prediction}} These studies aimed at predicting the performance metrics of a system, such as response time, resource utilization, and throughput, based on architecture models of the system; they employed architecture models to capture the abstract behaviors or component structures of a system, enriched with performance properties, such as resource and workload. The architecture models were then transformed into mathematical performance models, such as \textit{Queuing Networks} and \textit{Stochastic Petri Nets}, or simulation code to derive performance measures. Practitioners could benefit from such techniques by assessing performance and addressing concerns early in the development life cycle. 

\paragraph*{\textbf{Category 1.2 Retrospective Root Cause Analysis}} Four studies~\citeS{zhao2020butterfly, zhao2020performance, wu2020microrca} provided retrospective analysis of the root causes of performance issues. \citeS{zhao2020performance} revealed that a significant portion (around 30\%) of real-life performance issues require design-level optimization, i.e., simultaneously revising a group of related source code files.  This study used a new modeling technique, named \textit{Diff Design Space Matrix (D-DSM)}, to capture the essential design structure changes in code revisions for performance issues. Another study~\citeS{wu2020microrca} proposed an attributed graph model to investigate performance anomaly propagation across micro-services. \citeS{zhao2020butterfly} proposed a fine-grained (i.e. method-level) architectural modeling approach, named \textit{Butterfly Space} modeling, that combined the analysis of dynamic profiling metrics and static method-call dependencies to understand architectural patterns behind performance issues. 

\paragraph*{\textbf{Category 1.3 Model-based Architectural Optimization}} There are three studies~\citeS{du2015evolutionary, arcelli2019exploiting} in this sub-category which automatically recommend optimal architectural solutions by exploring the design space of a system. \citeS{du2015evolutionary} proposed an evolutionary algorithm for rule-based software performance optimization at the software architecture level. The proposed approach addressed the limited search space and uncertain application of rules in traditional rule-based performance optimization. Similarly, \citeS{arcelli2019exploiting} integrated performance monitoring data and architectural modelling to detect performance problems and suggest architectural changes.

Given the large volume of the studies in Category 1.1  we further mapped them into five clusters based on the focus of each study. Of particular note, one study~\citeS{martens2011monolithic} compared different model-based prediction methods in terms of the prediction accuracy and the required manual effort, for providing insights to practitioners. This paper was not classified into any of the five clusters described below:

\textbf{Category 1.1.1} contains studies that contributed prediction techniques for specific architecture styles. Nine studies focused on distributed architectures~\citeS{happe2014stateful, eshraghian2015performance, pinciroli2021model, didona2015enhancing, jindal2019performance}. They employed modeling techniques with rich notations for capturing the behavior and dependencies of  distributed system components, such as \textit{Directed Acyclic Graph (DAG)}~\citeS{eshraghian2015performance}, \textit{Discrete-Time Markov Chain (DTMC)}~\citeS{pinciroli2021model}. In particular, \citeS{eshraghian2015performance} focused on  service-oriented architectures. \citeS{gomez2014performance} focused on interoperable architecture, which enabled diverse interaction modes captured as performance parameters in the model. 
   
\textbf{Category 1.1.2} contains studies that targeted the performance prediction of specific systems. 
The most well studied were data intensive systems~\citeS{gribaudo2018performance, wu2015exploring, werle2020data, requeno2017performance, noorshams2014enriching, yasaweerasinghelage2018predicting, kross2017model, nalepa2015model, klinaku2021architecture}. 
Computational resources were critical due to the high demand for data storage, transmission, and manipulation. These studies all considered  parameters for representing resource demand in the modeling process. 


Three studies focused on performance prediction for systems featured by hardware and software co-design, including cyber-physical system~\citeS{pinciroli2021model}, real-time embedded system~\citeS{nguyen2017parad}, and Internet of Things system~\citeS{smith2017spe}. Hardware-related properties, such as communication delay among hardware components, were considered in the modeling process. In addition, architecture patterns for the hardware and software co-design are assessed for deriving the system performance~\citeS{pinciroli2021model, nguyen2017parad}. \citeS{maddodi2020aggregate} focused on performance prediction for systems with dynamic and varying processing requests.

\textbf{Category 1.1.3} contains studies that refined  common processes in  performance prediction techniques for automation, efficiency, and accuracy improvement~\citeS{gorsler2014performance, walter2017expandable, gomez2018enabling, eismann2019integrating, brosig2014quantitative, woodside2014transformation, etxeberria2014performance}. \citeS{walter2017expandable, gorsler2014performance} focused on developing automated methods for extracting and constructing architecture and performance models~\citeS{walter2017expandable,gomez2018enabling}.
\citeS{etxeberria2014performance} annotated performance properties and parameters on architecture models. In particular, practitioners often faced uncertain performance properties, such as operational profile and resource demand, which compromised the accuracy of performance prediction. \citeS{etxeberria2014performance} focused on explicitly capturing and managing such uncertainties in the modeling process. 
\citeS{eismann2019integrating} made the performance model solution and simulation faster, since the time required to solve or simulate a model increases exponentially with  system size. Finally, \citeS{gorsler2014performance} simplified and automated the overall prediction process, by proving a unified interface for performance queries, independent of the employed modeling formalism.
    
\textbf{Category 1.1.4} contains studies~\citeS{trubiani2020visarch, trubiani2015exploiting} that focused on providing traceable architectural feedback based on  performance prediction results. \citeS{trubiani2020visarch} focused on visualizing possible architectural refactorings based on  predicted results. \citeS{trubiani2015exploiting} focused on building backward traceability links from predicted results to the performance and architecture models to help in interpreting the prediction results. 
    
    
\textbf{Category 1.1.5} contains studies that enabled performance prediction under specific settings. Four studies~\citeS{arcelli2015control, brosig2014architecture, eismann2018modeling, huber2016model} enabled run-time or online performance prediction. This pushed the boundary of  performance prediction that traditionally only occurred at design time, to satisfy the needs of continuous system reconfiguration in modern  applications. These studies featured flexible architecture models, such as \textit{Descartes Modeling Language}~\citeS{eismann2018modeling, brosig2014architecture, huber2016model}, for annotating dynamic performance parameters in the architecture models. 3 studies~\citeS{mazkatli2020incremental, li2016evaluating ,brunnert2017continuous} contributed prediction techniques that allow continuous calibration of architecture model and performance parameters to reflect continuous system evolution.  
\subsubsection{\textbf{Purpose 2: Performance Anti-pattern Detection \& Resolution}}

The second high-level study purpose is focused on architectural anti-patterns that lead to performance problems. We identified 16 studies in this category~\citeS{chen2014detecting, arcelli2017applying ,fioravanti2017engineering,de2017model,arcelli2018performance,arcelli2015performance,trubiani2014exploring,trubiani2014guilt,cortellessa2014approach,trubiani2018exploiting, wert2014automatic, avritzer2021multivariate}. 
Performance anti-patterns are recurring issues in design or implementation that lead to negative performance consequences. The three studies by Smith \textit{et al.}~\cite{smith2000software, smith2002new, smith2003more} summarize 12 anti-patterns from  prior studies. Each performance anti-pattern is usually paired with a problem and its solution.
For example, \textit{Blob/God Class} was characterized by a high number of messages transmitted between a single big component and a large number of other components. It could be resolved by breaking down the responsibilities of the \textit{Blob/God Class} to smaller parts, to split the ``\textit{busy}" message traffic.

Related studies contributed automated approaches for detecting and refactoring such performance anti-patterns. The detection of anti-patterns usually relied on a set of rules that combine the analysis of performance metrics and architectural characteristics. The detection rules usually compared the performance metrics with a certain threshold, which were either obtained from \textcolor{black}{model-based prediction~\citeS{cortellessa2014approach, avritzer2021multivariate, trubiani2011detection, trubiani2014guilt, de2017model, arcelli2018performance, arcelli2015performance, trubiani2014exploring}} or dynamic profiling~\citeS{arcelli2017applying,fioravanti2017engineering,trubiani2018exploiting,wert2014automatic,chen2014detecting}.The detection rules also matched typical characteristics in the architecture model that formed anti-patterns. For example, the \textit{MessageSize Rule} checked if the deviation of service throughput associated with different components exceeds the respective threshold. Once matched, the \textit{MessageSize Rule} identified a potential bottleneck in the \textit{Pipe and Filter} pattern.

In addition to anti-pattern detection, 12 studies~\citeS{trubiani2014guilt, de2017model, arcelli2018performance, arcelli2015performance, trubiani2014exploring,trubiani2014exploring,fioravanti2017engineering,trubiani2018exploiting} provided resolutions to eliminate the anti-patterns. For example, ``\textit{changing scheduling algorithms to enable concurrent execution}'' resolves \textit{Concurrent Processing Systems}; and ``\textit{alleviate the congestion and use \textit{Shared Resource Principle} to minimize conflicts}'' resolves \textit{One-Lane Bridge}.

\subsubsection{\textbf{Purpose 3: Profiling \& Comparison}} 12 (11\%) studies focused on dynamically profiling and comparing the run-time performance of alternative architectural solutions for implementing the same applications~\citeS{avritzer2018quantitative, fioravanti2016experimental, haughian2016benchmarking, sunardi2019mvc, lung2014measuring, gonccalves2021monolith, alzboon2022performance, ngo2022evaluating, blinowski2022monolithic, avritzer2020scalability, hasselbring2020kieker}. 
\textcolor{black}{The architectural variables considered in these studies span a broad spectrum, including: 1) the types of data storage systems employed, such as \textit{Cassandra}, \textit{MongoDB}, \textit{NoSQL}, and \textit{MySQL}~\citeS{haughian2016benchmarking, fioravanti2016experimental}; 2) architectural patterns such as such as \textit{Leader/Follower} pattern and \textit{Half-Sync/Half-Async} pattern~\citeS{lung2014measuring} and \textit{Workload Smoother} pattern~\citeS{ngo2022evaluating}; 3) web-service architecture solutions~\citeS{sunardi2019mvc, alzboon2022performance}; and 4)  deployment architectures~\citeS{avritzer2018quantitative, gonccalves2021monolith, blinowski2022monolithic, avritzer2020scalability}.}

Alternative architectures for data storage we compared in ~\citeS{haughian2016benchmarking, fioravanti2016experimental}. \citeS{haughian2016benchmarking} compared the performance of two widely-used database management systems with different architectures, \textit{Cassandra} and \textit{MongoDB}. \textit{Cassandra} was for storing large amount of unstructured data. It had a multi-master architecture where data was distributed evenly across clusters to guarantee load balancing. \textit{MongoDB} was a document-oriented NoSQL data storage system that automatically partitioned data across multiple servers---known as sharding. It had a master-slave architecture where all writing operations must be directed to a single master cluster. The comparison showed that under heavy-reading workload, \textit{MongoDB} slightly outperformed \textit{Cassandra}. However, under heavy-writing workload, \textit{Cassandra} substnatially outperformed \textit{MongoDB}. Similarly, \citeS{fioravanti2016experimental} compared different implementations of a web application with three alternative databases---the traditional \textit{Hibernate + MySQL}, and two \textit{NoSQL} data storage systems, \textit{MongoDB} and \textit{Neo4j}. The comparison showed that both \textit{MongoDB} and \textit{Neo4j} outperformed \textit{MySQL + Hibernate} in scenarios of processing document-oriented and graph-oriented data.

\textcolor{black}{Two studies compared and contrasted various architectural patterns that can impact system performance.} \cite{lung2014measuring} compared two widely-used concurrent processing architectures, \textit{Half-Sync/Half-Async (HS/HA)} and \textit{Leader/Follower (LF)}. The \textit{HS/HA} pattern consisted of an asynchronous layer for the high priority requests in a message queue to avoid packet loss, and a synchronous layer for other requests and necessary I/O and computational operations. \textit{HS/HA} had the advantage of decoupling the asynchronous and synchronous services to avoid multi-thread blocking. In comparison, \textit{LF} had only one leader thread at all times, and when this thread finished processing a request, it became a follower thread and returned to the thread pool. The advantage of \textit{LF} was to minimize overhead and prevent race conditions. The experimental results showed that \textit{LF} performed slightly better than \textit{HS/HA} under a small workload; while \textit{HS/HA} performed much better with a large number of threads. 
\textcolor{black}{The study \citeS{ngo2022evaluating} evaluated the implementation of a workload smoother design pattern on Function as a Service (FaaS) cloud platforms, such as Amazon AWS Lambda, IBM, and Azure Cloud Function. A workload smoother  helps distribute traffic evenly over time, thus reducing bursts and ensuring smoother performance. The results indicated that different FaaS platforms adopted unique scaling strategies, and by applying a workload smoother, software engineers could achieve a success rate of 99 - 100\%, compared to a 60 - 80\% rate when the FaaS system was saturated.}

\textcolor{black}{The comparison of different web service architecture solutions has been the focus of the two studies~\citeS{sunardi2019mvc, alzboon2022performance}.} \citeS{sunardi2019mvc} explores two different MVC implementations - the \textit{Laravel Framework} and the \textit{Slim Framework}. The \textit{Laravel Framework}, being a full-stack MVC framework, enabled programmers to concentrate more on implementing business logic. Conversely, the \textit{Slim Framework} contained only a View, with the Controller and Models added as needed. They concluded that for small projects, the \textit{Slim Framework} was more suitable, whereas, for larger projects with numerous functions and libraries, the \textit{Laravel Framework} was the better choice. 
\textcolor{black}{\citeS{alzboon2022performance} compares the performance between 2-tier and 3-tier cloud applications, specifically Ghost and WordPress. Ghost, a blog cloud application without a database, was pitted against WordPress, a similar application but employing RDS (Relational Database Service) to host a Maria database. Their findings indicated that WordPress, a 3-tier blogging system, not only performed better but also exhibited greater stability than the 2-tier Ghost blogging system.}


Four studies~\citeS{avritzer2018quantitative, gonccalves2021monolith, blinowski2022monolithic, avritzer2020scalability} focused on comparing deployment architectures. \textcolor{black}{Four of these studies concentrated on profiling and comparing  deployment architectures, specifically between monolithic and micro-service architectures~\citeS{avritzer2018quantitative, gonccalves2021monolith, blinowski2022monolithic, avritzer2020scalability}.} \citeS{avritzer2018quantitative} compared micro-service architecture deployment environments, taking into account factors like memory allocation, CPU fraction in the deploying servers, and the number of Docker container replicas assigned to each micro-service. 
\textcolor{black}{\citeS{gonccalves2021monolith} delved into the migration of a large object-oriented system from a monolithic to a modular, microservice architecture, removing circular dependencies between modules. Performance metrics before and after the architecture migrations were  compared, revealing variations in processing times, throughput, and other parameters.}
\textcolor{black}{\citeS{blinowski2022monolithic}  contrasted the performance and scalability of monolithic and microservice architectures within a reference web application, concluding that a monolith excels in single-machine performance compared to a microservice-based architecture.}
\textcolor{black}{Lastly, \citeS{avritzer2020scalability} proposed a quantitative approach for the performance assessment of microservice deployment alternatives. Their four-step methodology incorporated operational profile data analysis, experiment generation, baseline requirements computation, and experiment execution, thereby providing a comprehensive lens to evaluate deployment configurations.}
\cite{zulkipli2013empirical} compared the performance of web applications deployed on two server architectures, i.e., 1) the \textit{Active/Active} environment that consisted of multiple redundant servers where the workload was balanced by a load-balancer installed between web services and web applications; and 2) \textit{Active/Passive} environment that consisted of a single server that provided service at any time while the other servers were used as standby devices. The study showed that the \textit{Active/Active} deployment architecture could provide more stable processing time and CPU utilization.

\subsubsection{\textbf{Purpose 4: Self-Adaption}}

Researchers have proposed a number of frameworks that monitor the performance of the system, and dynamically switch to the most efficient architectural solution based on run-time workload~\citeS{gergin2014decentralized, lung2016improving, camara2020quantitative, arcelli2020multi, ezzeddine2021design, peng2021parallel}.

The study \citeS{gergin2014decentralized} presented a decentralized, autonomic architecture for managing multi-tier transaction applications deployed on the cloud environment. For each tier in the application, an autonomic controller (in form of a \textit{Java} class) monitored the resource utilization of the tier, then enacted a set of elasticity policies that dynamically scale  the acquisition and release of cloud resources to and from the tier. The experiments demonstrated the effectiveness of the proposed architecture.

\citeS{lung2016improving} proposed a framework named \textit{SAFCA}, which contained three well-known concurrent thread processing architectures--- \textit{Dynamic-thread-creation (DTC)}, \textit{Half-Sync and Half-Async (HS/HA)} and \textit{Leader-Followers (LF)}. The framework automatically switches to the most efficient mechanism, based on adaptation policies. The framework activates \textit{HS/HA} during stable, continuous workload. When a sudden burst of requests is detected, the framework automatically switches to \textit{DTC}. When the system experiences failures, the framework switches to \textit{LF} because it ensures higher reliability. After the system recovers, the framework switches back to \textit{HS/HA}.

\textcolor{black}{\citeS{ezzeddine2021design} proposed a self-adaptive framework was proposed to optimize tail latency in event-driven microservices within large-scale cloud applications. This adaptive dual-mode autoscaling mechanism features: 1) a reactive mode that actively monitors event arrival rates and adjusts consumer microservice instances to respond to load changes promptly, and 2) a proactive mode that employs an autoregressive prediction model to foresee future event loads, thus enabling preemptive scaling. This proactive approach, which continuously learns online using an exponentially weighted recursive least squares algorithm, prevents  latency spikes by facilitating a seamless transition between modes once a configured accuracy level is achieved. This swift mode-switching capability enhances system adaptability to fluctuating loads, thereby minimizing latency. Experimentally, the framework demonstrated its effectiveness in maintaining tail latency Service Level Agreement (SLA) guarantees while operating with a minimum number of replicas.}

\textcolor{black}{In \citeS{peng2021parallel}, a performance enhancement in the assembly and optimization of parallel component programs is achieved by employing a set of distinct software agents that collaboratively implement four adaptive strategies. These strategies are: 1) Dynamically adjusting the parallelism of components, which adapts to the varying resource availability and load, thus ensuring efficient utilization of resources; 2) Altering data partitioning to balance the computational load across components, enhancing overall system throughput; 3) Enabling component migration, shifting load away from high-usage nodes to those with low load, thereby reducing potential performance bottlenecks; 4) Modifying implementation by switching between different versions of component implementation codes based on resource and performance requirements, thereby maintaining optimal system performance. The strategies' effectiveness was verified through experimentation on heterogeneous computer clusters, exhibiting significant performance benefits and flexibility over traditional methods.}

In recent years, researchers started to employ machine learning techniques in self-adaptive systems~\citeS{arcelli2020multi, camara2020quantitative}, rather than relying on pre-defined adaption policies or algorithms. 
Machine learning was used in making decisions about switching among alternative architectural solutions~\citeS{arcelli2020multi}, and in predicting the performance behavior of a system under execution environments~\citeS{camara2020quantitative}. \citeS{arcelli2020multi} exploited an evolutionary algorithm, the ``\textit{NSGA-II genetic algorithm}'', to suggest near-optimal alternative architectures in terms of mean response time for different system operational modes.  \citeS{camara2020quantitative} used a classic machine learning model,  ``\textit{Long-Short Term Memory (LSTM) Networks}'', to forecast the behavior of each component of a system to facilitate adaption. 
\subsubsection{\textbf{Facilitated Activities}}

\begin{table*}[h]
\setlength{\tabcolsep}{8pt}
\centering
\caption{Proportion of Research Purposes according to Development Activities (RQ1)}\label{Tab1}
\begin{tabular}{@{}lccccc}   
\hline
\textbf{Development Activity} &  \textbf{Design} & \textbf{Implementation} & \textbf{Deployment} & \textbf{Operation} & \textbf{Maintenance} \\[0.75ex] \hline\hline
1. Model-based Performance Analysis & \cellcolor[HTML]{656565}\color[HTML]{FFFFFF} 50 (48\%) & 4 (4\%) & 1 (1\%) & 8 (8\%) & 9 (9\%) \\ 
\indent{} 1.1. Performance Prediction & \cellcolor[HTML]{656565}\color[HTML]{FFFFFF} 46 (44\%) & 4 (4\%) & 1 (1\%) & 8 (8\%) & 6 (6\%) \\
\indent{} 1.2 Retrospective Root Cause Analysis & 0 &  0 & 0 & 0 & \cellcolor[HTML]{656565}\color[HTML]{FFFFFF} 4 (4\%) \\
\indent{} 1.3 Architectural Optimization & \cellcolor[HTML]{656565}\color[HTML]{FFFFFF} 3 (3\%) & 0 & 0 & 0 &  0  \\ \hline
2. Performance Anti-patterns Detection and Solution & \cellcolor[HTML]{656565}\color[HTML]{FFFFFF} 10 (10\%) & 1 (1\%) & 1 (1\%) & 0 & 5 (5\%) \\ \hline
3. Profiling and Comparison &  \cellcolor[HTML]{656565}\color[HTML]{FFFFFF} 4 (4\%) & 0 & \cellcolor[HTML]{656565}\color[HTML]{FFFFFF} 4 (4\%) & 2 (2\%) & 1 (1\%) \\ \hline
4. Self-Adaption & 1 (1\%) & 0 & 2 (2\%) & \cellcolor[HTML]{656565}\color[HTML]{FFFFFF} 6 (6\%) & 0 \\ \hline\hline

\textit{All} & \cellcolor[HTML]{656565}\color[HTML]{FFFFFF} 67 (64\%) & 6 (6\%) & 8 (8\%) & 16 (15\%) & 15 (14\%) \\ \hline
\end{tabular}
\label{tbl:life-cycle}
\end{table*}

We also the development activities that different studies aimed to facilitate, and maps the identified activities to the research objectives from RQ1. This provides guidance for practitioners to identify the most relevant techniques for the development activities they are engaged in.

Five software development activities (or phases) emerged from our analysis: 1) \textit{Software Design} when the developers focus on the high-level and preliminary architecture and design of their systems; 2) \textit{Software Implementation} when developers implement and test their systems; 3) \textit{Software Deployment} when developers launch their systems in the physical environment; 4) \textit{Software Operation} when developers focus on the run-time behavior of their systems in the product environment; and 5) \textit{Software Maintenance} when developers keep updating and supporting existing systems to solve arising issues.

In Table~\ref{tbl:life-cycle},  rows 2 to 5 show the number and percentage of studies in each purpose that addressed each development phase. The cells colored grey highlight the largest numbers in each  category. The bottom row in Table~\ref{tbl:life-cycle}, (``\textit{All}''), shows the total number (and percentage) of studies for each phase. Next we  elaborate the details of each study purpose.

\paragraph{\textbf{Purpose 1: Model-based Performance Analysis:}} 
In \textit{Category-1.1 Performance Prediction}, most studies  (\textcolor{red}{XX\%}) 
aimed at facilitating software design ~\citeS{happe2014stateful, gomez2014performance, eshraghian2015performance, smith2017spe, requeno2017performance, kross2017model, wu2015exploring, gribaudo2018performance, maddodi2020aggregate, werle2020data, noorshams2014enriching, yasaweerasinghelage2018predicting, nalepa2015model, nguyen2017parad, brosig2014quantitative, eismann2019integrating, gomez2018enabling, woodside2014transformation, gorsler2014performance, trubiani2020visarch, pinciroli2021model, klinaku2021architecture}. 
These studies conducted  early performance assessments based on  design models. The performance metrics predicted based on these models helps to evaluate early design choices, to avoid  rework before a system is implemented. 
Eight (13\%) studies targeted run-time performance prediction~\citeS{mazkatli2020incremental,arcelli2015control,brosig2014architecture,eismann2018modeling,huber2016model}. They calibrated performance models with resource usage profiles while systems were running. Six (10\%) studies targeted software maintenance~\citeS{brunnert2017continuous,li2016evaluating}. They derived software models from existing systems and provided performance evaluation considering architectural evolution. 

The four studies in Category-1.2 Retrospective Root Cause Analysis~\citeS{wu2020microrca, zhao2020butterfly, zhao2020performance}  target the software maintenance phase. They perform a root cause analysis of performance issues in existing systems, providing  guidance for developers doing maintenance. 

All three studies in \textit{Category-1.3 Model-based Architectural Optimization} aimed at facilitating better software design solutions~\citeS{arcelli2019exploiting,du2015evolutionary}. They 
provided support for stakeholders in performing architectural refactoring on design models.

\paragraph{\textbf{Purpose 2: Performance Anti-pattern Detection and Resolution}} 

The majority of the studies in \textit{Purpose-2} (ten out of sixteen, or 63\%)~\citeS{cortellessa2014approach,trubiani2014guilt,de2017model,arcelli2018performance,arcelli2015performance,trubiani2014exploring,arcelli2017applying} focus on the identification and refactoring of performance anti-patterns during the software design phase. These studies  employ UML diagrams and \textit{PCM} as their analytical tools.

Conversely, five studies (31\%)~\citeS{fioravanti2017engineering, chen2014detecting, trubiani2018exploiting, wert2014automatic} emphasize the maintenance phase of software development. They apply dynamic profiling to existing software applications and use performance metrics to identify and eliminate anti-patterns, thus helping maintain the health and efficiency of the systems.
Additionally,  paper~\citeS{avritzer2021multivariate} extends this scope to include the implementation and deployment phases. This study introduces a novel approach for Software Performance Anti-pattern (SPA) characterization and detection,  designed to bolster continuous integration/delivery/deployment (CI/CD) pipelines. This inclusion addresses the computational efficiency gaps in current detection algorithms,  enhancing performance across all development stages.

\paragraph{\textbf{Purpose 3: Profiling and Comparison}}

In this category, four (36\%) studies,  \citeS{haughian2016benchmarking, fioravanti2016experimental, sunardi2019mvc, lung2014measuring}, concentrated on the software design phase by providing empirical experience regarding design choices like databases, web frameworks, and programming paradigms. Focusing on the deployment phase, four (36\%) studies, \citeS{avritzer2018quantitative, alzboon2022performance, avritzer2020scalability}, offer quantitative assessment and guidance for deployment decisions, including architectural comparisons and performance assessment of  deployment configurations in micro-services and web-services environments. Two (18\%) studies, \citeS{ngo2022evaluating, blinowski2022monolithic}, emphasized the operation phase by exploring the effectiveness of workload smoother  patterns and comparing the performance and scalability of different architectural styles in operational conditions. One study, \citeS{gonccalves2021monolith}, with a  focus on the maintenance phase, presented a detailed refactoring process, from a large monolithic to a modular system, analyzing the performance impact and refactoring effort.

\paragraph{\textbf{Purpose 4: Self-Adaption}} 
Self-adaptive systems dynamically adjust their configuration at run-time to meet performance goals in changing environments. The majority of studies (77\%) focused on system operation at run-time~\citeS{lung2016improving, camara2020quantitative, gergin2014decentralized, arcelli2020multi, ezzeddine2021design}. Notably, two studies, \citeS{gergin2014decentralized} and \citeS{peng2021parallel}, went a step further and not only reconfigured the software components but also re-deployed resources in the deployment environment. Consequently, these studies also offer valuable insights for the deployment phase.

\textbf{RQ1 Summary:} The 109 studies that integrate software architecture and performance analysis serve four objectives 1) model-based performance analysis; 2) performance anti-pattern detection and resolution; 3) profiling and comparison of architectural alternatives; and 4) self-adaptive architecture for dynamic performance optimization. These studies tackle challenges covering five major phases of the software development life-cycle, namely \textit{Design}, \textit{Implementation}, \textit{Deployment}, \textit{Operation}, and \textit{Maintenance}. Overall, \textit{Design} is the most extensively covered.

\subsection{RQ-2: Classic Study Templates}
\label{sec:rq2}

RQ2 provides ``cheat-sheets'' for practitioners to quickly familiarize themselves with study templates that integrate architecture and performance analysis. This can help them quickly get started if they are interested in replicating or extending existing techniques. This RQ reveals how software architecture and performance analyses are combined for different study purposes. It offers an in-depth understanding of: 1) which architecture models are used and how they are analyzed; 2) which performance models or metrics are used, and how they are collected and analyzed; and most importantly, 3) how the architecture and performance analysis are integrated for addressing performance concerns. We summarize typical study templates for different study purposes, which reveal the common study processes. To further facilitate comprehension, we've color-coded the common steps across the templates, which visually highlights their shared elements and helps practitioners quickly discern the commonalities and differences among them.

\subsubsection{\textbf{Model-based Prediction Template}} \label{sec:template-p1}
This section presents the study template summarized from sub-category 1.1, \textit{model-based performance prediction}, which contains the majority (70\%) of the studies in the model-based analysis. It is not meaningful to summarize a study template for the other two sub-categories (1.2 and 1.3) with only three studies each.

\begin{figure*}[h]
    \centering
    \begin{subfigure}{.5\textwidth}
        \centering
        \includegraphics[width=.95\linewidth]{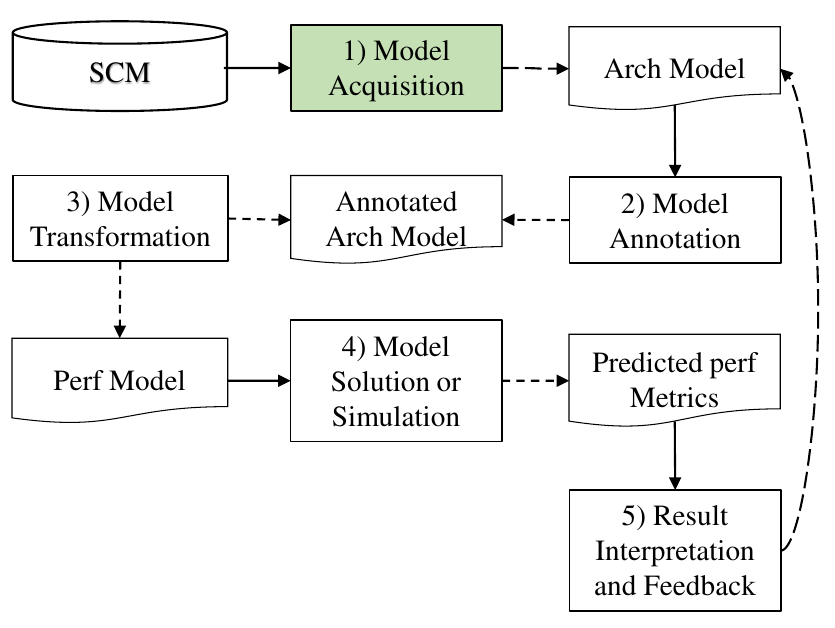}
        \caption{Model-based Prediction Template}
        \label{fig:template-p1}
    \end{subfigure}%
    \begin{subfigure}{.5\textwidth}
        \centering
        \includegraphics[width=.95\linewidth]{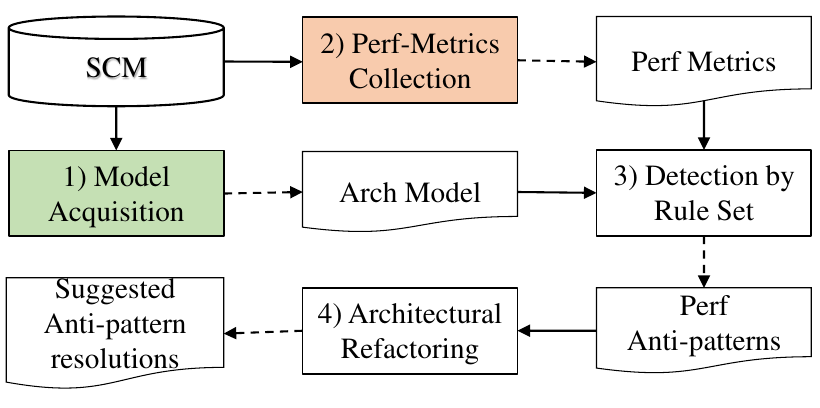}
        \caption{Anti-pattern D\&R Template}
        \label{fig:template-p2}
    \end{subfigure}
    \begin{subfigure}{.5\textwidth}
        \centering
        \includegraphics[width=.95\linewidth]{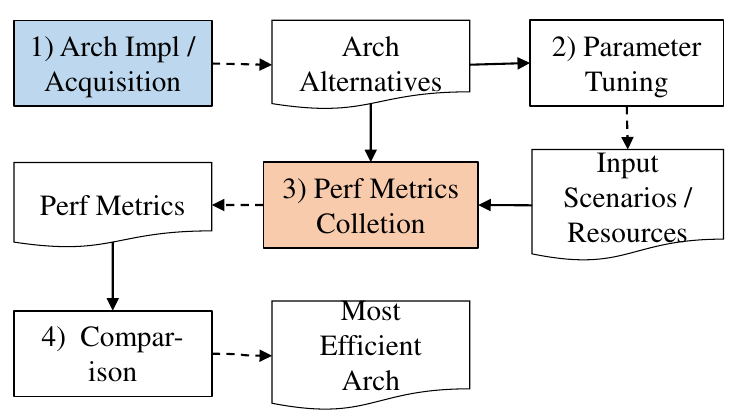}
        \caption{Profiling and Comparison Template}
        \label{fig:template-p3}
    \end{subfigure}%
    \begin{subfigure}{.5\textwidth}
        \centering
        \includegraphics[width=.95\linewidth]{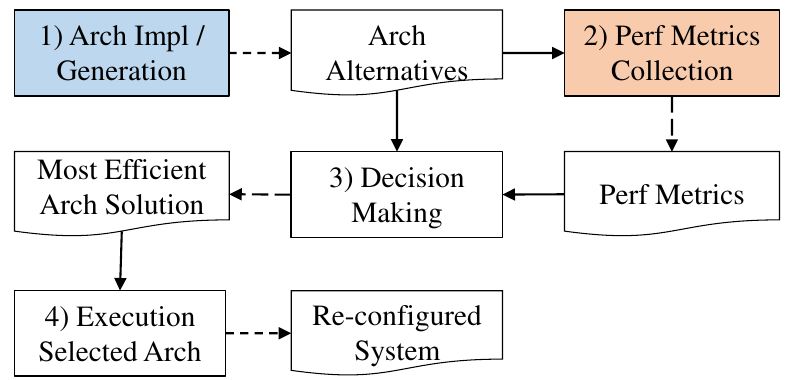}
        \caption{Self-Adaption}
        \label{fig:template-p4}
    \end{subfigure}
    \caption{Classic Study Templates}
    \label{fig:study-template}
\end{figure*}

As depicted in Figure~\ref{fig:template-p1}, the model-based performance prediction studies encompass five phases: 1) Model Acquisition, in which architecture models of a system are procured; 2) Model Annotation, involving the annotation of performance-related information on the architecture models; 3) Model Transformation, which transforms the annotated architecture models into executable performance models; 4) Model Solution or Simulation, that involves solving or simulating the performance models to predict performance metrics; and 5) Result Interpretation and Feedback, providing interpretation of the predicted results and feedback to stakeholders.


Note that, not every study fits 100\% to this template, which captures the major  steps we observed from the studies. In particular, most studies stopped at Step 4, i.e., obtaining the performance metrics. Among them, six  studies~\citeS{wu2015exploring, li2016evaluating, wu2020microrca, pinciroli2021model} did not have Step 3 since they used graph-based architectural models which do not need transformation before solution or simulation. Nine studies~\citeS{gomez2014performance, trubiani2015exploiting, maddodi2020aggregate, trubiani2020visarch, happe2014stateful} extended the process with Step 5, which traced back from the predicted results for providing interpretation of the prediction results and architectural feedback for stakeholders.

\textbf{Step 1: Model Acquisition:} We observed two methods to acquire a system model: 1) retrieving from project documentation and 2) recovering  via reverse-engineering (introduced in Section~\ref{sec:background}). 
Most of the  studies used the first method, where the architecture models were created in the design phase by the project team. Six studies~\citeS{walter2017expandable,brunnert2017continuous,li2016evaluating} used the second method.

We summarized three types of architecture models: 1) \textit{Behavioral Models (BM)} capture system architecture based on the collaborations among system objects, their internal state changes, the interactions, e.g., \textit{UML Sequence Diagrams},  \textit{UML Activity Diagrams},  \textit{UML Use Case Diagrams}, and  \textit{Business Process Execution Language (BPEL)} are the behavioral models. 2) \textit{ Component Models (CM)} capture the architecture of a system emphasizing the static characteristics as components, connectors, and their compositions, such as \textit{Palladio Component Model (PCM)}, \textit{Descartes Modeling Language (DML)}, \textit{Method Call Graph}, \textit{UML Class Diagram}, \textit{UML Component Diagram}, and \textit{UML Deployment Diagram} are used for this purpose. And 3) \textit{Hybrid Models (HM)} capture the system architecture as combinations of  component structure and behavioral interactions, including the \textit{Web Modeling Language (WebML)}, the \textit{Service-Oriented Architecture Modeling Language (SoaML)}, and \textit{Architectural Description Languages (ADLs)} are hybrid architecture models



\textbf{Step 2: Model Annotation: } The second step is to annotate the architecture models with performance-related information in preparation for performance prediction. 
We summarized five types of annotation data from the literature, including: 1) \textit{Workload}, which describes the run-time workload on a system during execution. The workload can be specified as a status or the intensity in numerical values. 2) \textit{Resource Demand}, which is the estimation of the resource demand for jobs, such as memory, CPU, and disks. 3) \textit{Probability}, which is the probability of executing a specific conditional process in a system. Processes with high probability have greater impact on performance than low probability processes. 4) \textit{Timing Specification}, which is the timing behavior of a system, such as the time intervals between requests, the time to acquire and release a job. And 5) \textit{Scheduling Policy}---the policy for scheduling workload requests, such as ``\textit{First In First Out (FIFO)}'', ``\textit{Last In First Out (LIFO)}'', and ``\textit{Processor Sharing (PS)}'', etc. 





\begin{table*}[h!]
\caption{Summary of Annotated Information with Applied Architecture Models}
\begin{adjustbox}{width=1.95\columnwidth,center}
\centering
\begin{tabulary}{\textwidth}{@{}*2{|L}*2{|L}*2{|L}*2{|L}*2{L}*2{L}}
 \hline
 \multicolumn{1}{>{\normalsize}c}{Model} &  \multicolumn{1}{>{\normalsize}c}{Workload} & \multicolumn{1}{>{\normalsize}c}{Resource Demand}  & \multicolumn{1}{>{\normalsize}c}{Probability} &  \multicolumn{1}{>{\normalsize}c}{Timing Specification} &  \multicolumn{1}{>{\normalsize}c}{Scheduling Policy}  \\ [0.5ex] 
 \hline\hline
UML Diagrams  & \citeS{tribastone2010performance,wrzosk2012applying,perez2010performance,gomez2014performance,geetha2011framework,litoiu2013performance,maddodi2020aggregate,berardinelli2010performance,woodside2014transformation,trubiani2013model,gomez2018enabling,alhaj2013traceability,arcelli2013software,eramo2012performance,arcelli2015control} & \citeS{tribastone2010performance,wrzosk2012applying,perez2010performance,khan2010performance,gomez2014performance,geetha2011framework,requeno2017performance,berardinelli2010performance,etxeberria2014performance,tawhid2011automatic,woodside2014transformation,trubiani2013model,arcelli2013software,arcelli2015control} & \citeS{etxeberria2014performance,eramo2012performance,arcelli2015control} & \citeS{gomez2014performance,litoiu2013performance,etxeberria2014performance,tawhid2011automatic,trubiani2013model} & \citeS{khan2010performance,requeno2017performance,eramo2012performance} \\  \hline

PCM & \citeS{eshraghian2015performance,werle2020data,yasaweerasinghelage2018predicting,kross2017model,noorshams2014enriching,brosig2014quantitative,von2011performance,brunnert2017continuous} & \citeS{happe2014stateful,werle2020data,yasaweerasinghelage2018predicting,kross2017model,happe2010parametric,hauck2011ginpex,hauck2010automatic,groenda2012improving,brosig2014quantitative,brosig2011automated,von2011performance,brunnert2017continuous} & \citeS{happe2014stateful,happe2010parametric,brosig2014quantitative,brosig2011automated} & \citeS{eshraghian2015performance} & \citeS{kross2017model,hauck2011ginpex,hauck2010automatic} \\  \hline

DML & \citeS{walter2017expandable,brosig2014architecture} & \citeS{eismann2019integrating,walter2017expandable,brosig2014architecture,eismann2018modeling} & \citeS{eismann2019integrating,brosig2014architecture,eismann2018modeling} &  &  \\  \hline

Calling Dependency Graph & \citeS{jalali2011new,nalepa2015model,hill2010run} & \citeS{wu2015exploring,nalepa2015model,hill2010run} & \citeS{jalali2011new} &  & \citeS{wu2015exploring,gribaudo2018performance} \\  \hline

BPEL & \citeS{wrzosk2012applying} & &  & \citeS{wrzosk2012applying} & \\  \hline 

 WebML &  & \citeS{gambi2010model} & \citeS{gambi2010model} & &  \\  \hline

 SoaML  & \citeS{tribastone2010performance,mani2013propagation} & \citeS{mani2013propagation} & \citeS{mani2013propagation} & \citeS{mani2013propagation} & \\  \hline

ADLs  & \citeS{nguyen2017parad} & \citeS{nguyen2017parad,gaudel2011ada} &  &  & \citeS{nguyen2017parad} \\ \hline
\end{tabulary}
\end{adjustbox}
\label{tbl:arch-model-annotation}

\end{table*}

\label{sec:rq3-p1-step3}

\textbf{Step 3: Model Transformation: } The third  step is to transform an architecture model, annotated with performance related parameters, to a form that can be used to predict  performance metrics. We observed two types of transformation methodologies---\textit{Model-to-Model (M2M) Transformation} and \textit{Model-to-Text (M2T) Transformation}.

In \textit{M2M Transformation}, the  architecture model is transformed to a performance model, which  captures the dynamics of a system in operation. We found three types of performance models in the studies, including \textit{Queuing Networks} with four variations, \textit{Stochastic Petri Nets} with two variations, and \textit{Stochastic Process Algebras}. \textit{Queuing Networks} are composed of \textit{service centers} representing system resources and \textit{jobs} representing the consumers of the resources~\citeS{wu2015exploring}. \textit{Jobs} travel through a network of \textit{service centers} following probabilistic routes. This can be used to estimate the response time of the system by considering  waiting times, queue lengths, and server utilization. \textit{Stochastic Petri Nets} are composed of two types of elements: \textit{places} and  \textit{transitions}~\cite{requeno2017performance}. A \textit{place} can contain any number of tokens, which enables a \textit{transition} if all other \textit{places} connected to it contain at least one token. The transitions fire after a probabilistic delay determined by a random variable. \textit{Stochastic Petri Net} are often used to model step-wise processes in distributed systems, which include choice, iteration, and concurrent execution~\citeS{pinciroli2021model, requeno2017performance}. Finally, \textit{Stochastic Process Algebras} are derived from classical process algebras, which are abstract languages used for the specification  of concurrent systems~\citeS{eshraghian2015performance}. Systems are modelled as collections of entities, called agents, which execute atomic actions. The atomic actions describe the behavior of the entities in concurrent execution.

The \textit{M2M Transformation} is usually based on graph transformation. There are third-party tools to automate the process, including, \textit{PUMA}~\citeS{woodside2014transformation}, \textit{PRIMA-UML}~\citeS{cortellessa2014approach}, \textit{ArgoSPE}~\citeS{gomez2014performance}, \textit{KLAPER}~\citeS{gorsler2014performance}, \textit{TwoTower}~\citeS{de2017model}, and \textit{S-PMIF+}~\citeS{gomez2018enabling, smith2017spe}. Some architecture models have default model transformation techniques, such as \textit{PCM2LQN}~\citeS{brosig2014quantitative} and \textit{PCM2QPN}~\citeS{brosig2014quantitative} for the \textit{PCM} model, and \textit{DML2LQN}~\citeS{huber2016model} for \textit{DML} model.
Some studies use declarative model transformation languages, including \textit{Query/View/Transformation (QVT) language}~\citeS{gomez2018enabling}, \textit{Janus Transformation Language (JTL)}~\citeS{arcelli2019exploiting}, and \textit{Gra2Mol}~\citeS{de2017model}. A few studies leverage meta-models to improve the accuracy while mapping components in the source model to the entities in the target model, including \textit{Core Scenario Model (CSM)}~\citeS{woodside2014transformation} and \textit{Graph Transformation System (GTS)}~\citeS{eshraghian2015performance}.

In \textit{M2T Transformation}, an input architecture model is transformed to a textual artefact, which serves as executable simulation code~\cite{becker2008coupled}. The most commonly used \textit{M2T Transformation} technique is \textit{PCM2SimuCom}, which translates \textit{PCM} into executable simulation code~\citeS{brosig2014quantitative,noorshams2014enriching,brunnert2017continuous,yasaweerasinghelage2018predicting,kross2017model,trubiani2014guilt,happe2014stateful}. \textit{SimuCom} is the default analysis tool for \textit{PCM}. It packs the translation and execution function of simulation code together. Similarly, \textit{SimuLizar}~\citeS{walter2017expandable} is a simulation tool for \textit{PCM}. In addition, the tools \textit{SLAstic.SIM} and \textit{SimSo} generate simulation code for \textit{Business Process Execulation Language (BPEL)} and \textit{ADLs} respectively.

\textbf{Step 4: Model Solution or Simulation:}
This step obtains performance metrics, such as response time, resource utilization, and system throughput, by  mathematically solving the performance model or executing the simulation code generated from the previous step. The advantage of simulation is its high accuracy, but it is  time consuming, since the simulation  can be  complicated when evaluating real-world systems~\cite{becker2008coupled, balsamo2004experimenting}. In comparison, solving the performance models is faster but less accurate. 

Performance models usually have  analytical solvers or simulators. For example, \textit{Queuing Networks} have multiple analytical solvers, such as \textit{LQN Solver (LQNS)}~\citeS{trubiani2020visarch, maddodi2020aggregate}, \textit{JMT Solver}~\citeS{etxeberria2014performance, trubiani2015exploiting}. Similarly, \textit{Petri Nets} has the analytical solver \textit{GreatSPN}~\citeS{requeno2017performance, gomez2014performance, woodside2014transformation} and the simulator \textit{SimQPN}~\citeS{gorsler2014performance,eismann2019integrating,huber2016model,walter2017expandable}. In comparison, the simulation code imitates the execution of a system over a period of time~\citeS{werle2020data} to obtain  performance metrics. The most commonly used simulation tool is \textit{SimuCom}~\citeS{happe2014stateful,kross2017model,yasaweerasinghelage2018predicting,noorshams2014enriching,brunnert2017continuous}, which focuses on executing the simulation code from \textit{PCM}. Another simulation tool, \textit{SimuLizar}~\citeS{walter2017expandable}, is more advanced than \textit{SimuCom}, with higher efficiency and reliability.

There are two model resolution techniques: analytical techniques
and simulation techniques
The analytical resolution technique addresses performance models generated by \textit{Model-to-Model} transformation. Analytical techniques are faster but less accurate than simulation. Different performance models usually have their own analytical resolution tools~\citeS{pinciroli2021model,barbierato2012perfbpel,requeno2017performance,gomez2014performance,smith2017spe,wrzosk2012applying,gambi2010model,huber2016model}. Simulation is the imitation of the operations of a system over a period of time~\citeS{werle2020data}.

\textbf{Step 5: Result Interpretation and Feedback:} As  mentioned earlier, nine studies contain a step  to interpret the prediction results and provide feedback to stakeholders~\citeS{gomez2014performance, trubiani2015exploiting, maddodi2020aggregate, trubiani2020visarch, happe2014stateful}.  \citeS{gomez2014performance, trubiani2015exploiting} focused on locating performance anomalies in architecture models based on  predicted metrics. They detected anomalies by comparing the predictions---response time~\citeS{trubiani2015exploiting}, resource utilization~\citeS{gomez2014performance, trubiani2015exploiting}, and throughput~\citeS{trubiani2015exploiting}---with pre-defined performance specifications. \citeS{maddodi2020aggregate, happe2014stateful} mapped the fluctuation in the predicted performance metrics with the respective workload scenarios to understand problems behind the prediction results. \citeS{trubiani2020visarch} visualized possible architectural refactorings based on the predicted results.

\subsubsection{\textbf{Performance Anti-Patterns D\&R Template}} 
Figure~\ref{fig:template-p2} illustrates the typical workflow of studies for \textit{Purpose-2: Performance Anti-pattern Detection and Solution}. There are four steps: 1) Model Acquisition, which acquires the architecture model of a system; 2) Performance Metrics Collection, based on either dynamic profiling or using model-based prediction; 3) Detection by Rule Set, which detects performance anti-patterns through a set of detection rules that combine the architecture model and performance metrics; and 4) Refactoring, which suggests refactoring to resolve the identified anti-patterns. Of note, two studies in \textit{Purpose-2} do not have the fourth step, since they only detect performance anti-patterns without providing solutions~\citeS{cortellessa2014approach, wert2014automatic}. 

\textbf{Step 1: Model Acquisition:}
The first step is to acquire the architecture model of the system under investigation. 9 related studies retrieved architecture models, such as UML diagrams, \textit{PCM}, and \textit{ADL}, from project designs~\citeS{arcelli2017applying, cortellessa2014approach, arcelli2015performance, arcelli2012antipattern, trubiani2014exploring, de2017model, arcelli2018performance, trubiani2014guilt, avritzer2021multivariate}. The other five studies~\citeS{fioravanti2017engineering, trubiani2018exploiting, chen2014detecting} recovered architecture models by reverse engineering existing software systems.

\textbf{Step 2: Performance Metrics Collection:} This step collects performance metrics by either dynamic profiling or model-based prediction. Six (40\%) studies used performance profiling tools, such as \textit{JMeter}~\citeS{trubiani2018exploiting}, \textit{YourKit}~\citeS{trubiani2018exploiting}, \textit{Dynamic Spotter}~\citeS{wert2014automatic}, or self-built profiling programs~\citeS{fioravanti2017engineering, arcelli2017applying}, to collect performance metrics, including response time~\citeS{trubiani2018exploiting, wert2014automatic, fioravanti2017engineering, trubiani2014exploring, de2017model, arcelli2018performance, cortellessa2014approach, arcelli2015performance, arcelli2017applying, trubiani2014guilt}, resource utilization~\citeS{trubiani2018exploiting, wert2014automatic, trubiani2014exploring, de2017model, arcelli2018performance, cortellessa2014approach, arcelli2015performance, trubiani2014guilt}, throughput~\citeS{trubiani2018exploiting, wert2014automatic, de2017model, arcelli2018performance, cortellessa2014approach, arcelli2015performance, trubiani2014guilt}, latency~\citeS{trubiani2018exploiting}, and data transmission loss ratio. Several studies also collected some metrics relevant to anti-pattern detection, such as the number of calls of a function~\cite{wert2014automatic}, the queue length of requests, and the size of transferred messages among system components. The remaining studies~\citeS{de2017model, arcelli2018performance, cortellessa2014approach, arcelli2015performance} collected performance metrics using model-based prediction methods. These studies followed the process summarized in template-1 (see Section~\ref{sec:template-p1}) to transform the architecture model into performance models, \textit{Queuing Network}~\citeS{trubiani2014exploring, arcelli2018performance, cortellessa2014approach, arcelli2015performance}, or \textcolor{black}{executable simulation code~\citeS{trubiani2014guilt, de2017model, avritzer2021multivariate}}.

The collected metrics (profiled or predicted) were used in the same way in follow-up steps. But the anti-patterns may impact how extensively the performance metrics were collected. The detection of \textit{Blob / God Class}, \textit{Pipe and Filter}, \textit{Concurrent Processing Systems}, \textit{Extensive Processing}, \textit{Circuitous Treasure Hunt}, \textit{Empty Semi Trucks}, \textit{One-Lane Bridge}, and \textit{Excessive Dynamic Allocation}, relied on sampling performance metrics during a period of time. That is, researchers only need to sample the average, maximal, or minimal value of related performance metrics during the profiling period~\citeS{wert2014automatic, trubiani2011detection, arcelli2012antipattern}. In comparison, the detection of \textit{Traffic Jam}, \textit{The Ramp}, and \textit{More is Less} relied on collecting the trend and fluctuation of the performance metrics over time{~\cite{trubiani2011detection}}.

\textbf{Step 3: Detection by Rule Set:} The detection of an anti-pattern usually relied on the combination of multiple rules to match both the dynamic metrics and architectural features~\citeS{wert2014automatic, cortellessa2014approach}.

\begin{table*}
\small
\setlength{\tabcolsep}{12pt}
\caption{Performance Anti-pattern Detection Rules}
\begin{tabulary}{\textwidth}{@{}*4{L}}
 \hline
Measurement-based Rule  & Description  \\ [0.5ex] 
 \hline\hline
 \textit{MessageSize Rule} & Avg size of message transmitted among a component $<$= Threshold   \\ 
\textit{Interaction Rule} & Number of external calls of a component $>$= Threshold   \\
\textit{Utilization Rule} & Resource (e.g. CPU, hard disk, memory) $>$= Threshold   \\ 
\textit{ServiceTh Rule} & Deviation of service throughput associated with different components $>$= Threshold  \\
\textit{WaitingTime Rule} & (1) Avg response time $>$= Threshold; (2) Response time increases/decreases continuously. \\ 
\textit{QueueLength Rule} & Queue length of requests $>$= Threshold  \\
\textit{Unbalanced Rule} & Deviation of a resource (e.g. CPU) utilization with different components $>$= Threshold \\
\textit{DiskMoreUtilized Rule} & (Hard disk utilization - CPU utilization) $>$= Threshold  \\  \hline
Model-based Rule & Description  \\ [0.5ex] \hline \hline
\textit{RemoteCommunication Rule} & Number of communications between local and remote components $>$= Threshold  \\
\textit{Structural Rule} & Scheduling policy of a component is a specific one, e.g. FIFO, LIFO, etc. \\
\textit{Usage Rule} & Number of dependencies of a component with components $>$= Threshold\\
\textit{Probability Rule} & Deviation of the probabilities of executing different processes in a component $>$= Threshold \\
 [1ex] 
 \hline
\end{tabulary}
\label{tbl:detecting-rules}
\end{table*}

Table~\ref{tbl:detecting-rules} lists the detection rules  from the literature. These rules are of two types: 1) rules based on on performance metrics (row 1 to  8), and 2) rules based on architecture features (row 9 to  12). Most rules work by comparing a  metric to a  threshold. For example, if the average size of transmitted message with a component is less than a  threshold, the \textit{MessageSize Rule} is matched. Some rules work by checking the \textit{deviation} of the  metric or feature over time or associated with different components. For instance, the \textit{ServiceTh Rule} checks the deviation of service throughput associated with  components to identify whether there is a bottleneck in a \textit{Pipe and Filter} architecture. The \textit{DiskMoreUtilized Rule} checks the difference between disk utilization and CPU usage.

Table~\ref{tbl:pattern-vs-rules} illustrates the mapping between each anti-pattern and its  detection rule(s). Note that four anti-patterns, \textit{Tower of Babel}, \textit{Excessive Dynamic Allocation}, \textit{The Ramp}, and \textit{More is Less}~\cite{smith2000software, smith2002new, smith2003more}, cannot be automatically detected by any rule yet. The other  anti-patterns can be automatically detected by  one or more rules{~\cite{wert2014automatic}}. For example, \textit{Blob/God Class} can be detected by matching three rules: 1) \textit{\textit{Usage Rule}:} a component has a dependency on a large number of other components; 2) \textit{Interaction Rule:} the component is called by a large number of external requests; and 3) \textit{Utilization Rule:} the component has high memory utilization{~\cite{wert2014automatic}.}

\begin{table}[t]
\setlength{\tabcolsep}{12pt}
\caption{Performance Anti-patterns vs. Detection Rules}
\begin{adjustbox}{width= \columnwidth,center}
\centering
\begin{tabulary}{\textwidth}{|p{3cm}|p{8cm}|}
 \hline
 \multicolumn{1}{|>{\large}c|}{Perf Anti-pattern} & \multicolumn{1}{|>{\large}c|}{Detection Rule} \\ [0.5ex] 
 \hline
\textit{Blob / God Class} & \textit{Interaction Rule}, \textit{Utilization Rule}, \textit{Usage Rule} \\ \hline
\textit{Concurrent Processing Systems} & \textit{Utilization Rule}, \textit{Unbalanced Rule} \\ \hline
\textit{Pipe and Filter} & \textit{ServiceTh Rule} \\ \hline
\textit{Extensive Processing} & \textit{Utilization Rule}, \textit{Structural Rule}, \textit{Probability Rule} \\ \hline
\textit{Circuitous Treasure Hunt} & \textit{Interaction Rule}, \textit{Message Size Rule}, \textit{DiskMoreUtilized Rule} \\ \hline
\textit{Empty Semi Truck} & \textit{Message Size Rule}, \textit{Interaction Rule}, \textit{Remote Communication Rule} \\ \hline
\textit{One-Lane Bridge} & \textit{Wating Time Rule}, \textit{QueueLength Rule} \\ \hline
\textit{Traffic Jam} & \textit{Waiting Time Rule} \\ \hline
\end{tabulary}
\end{adjustbox}
\label{tbl:pattern-vs-rules}

\end{table}

\textbf{Step 4: Architectural Refactoring:} Some studies further contributed refactoring solutions to fix the  anti-patterns, automatically or with manual effort~\citeS{trubiani2018exploiting,chen2014detecting,fioravanti2017engineering,trubiani2014exploring,de2017model,arcelli2018performance,arcelli2015performance,arcelli2017applying,trubiani2014guilt}. 

Three performance anti-patterns---\textit{Concurrent Processing Systems}, \textit{Extensive Processing}, and \textit{One-Lane Bridge}---can be automatically refactored. \textit{Concurrent Processing Systems} can be fixed by ensuring a balanced request distribution---i.e. through modifying the probability of branch actions in the problematic components~\citeS{arcelli2018performance}. \textit{Extensive Processing} can be resolved by using a different scheduling policy that grants short-running requests higher priority. \textit{One-Lane Bridge} can be refactored by optimizing the processing paths to alleviate the congestion of requests~\citeS{trubiani2014guilt}. 

The refactoring of \textit{three} anti-patterns, \textit{Blob / God Class}, \textit{Circuitous Treasure Hunt}, and \textit{Empty Semi Trucks}, are ``semi-automated''. The resolutions of these performance anti-patterns requires expert knowledge for optimizing the logic in the impacted components. The \textit{Blob / God Class} anti-pattern can be resolved by 1) manually optimizing the ``Blob / God Class'' component by delegating its business logic to the components that it communicates with, and 2) decreasing the number of calls to the \textit{Blob /God Class} component.  The \textit{Circuitous Treasure Hunt} anti-pattern can be resolved by 1) decreasing excessive communications with the database, and 2) manually restructuring the database tables to decrease excessive communication. 

For the other anti-patterns, we did not find refactoring solutions in the literature. The refactoring of these  anti-patterns requires changing the code  of the impacted components. This requires case-by-case analysis by an expert. For example, the refactoring of \textit{Pipe and Filter} and \textit{Towel of Babel} anti-patterns requires manual optimization of the business logic of the filter components and the translation component. The refactoring of the \textit{Ramp} anti-pattern requires that developers choose a more efficient data structure to process large amounts of data. The solutions to \textit{Traffic Jam} and \textit{More is Less} anti-patterns need developers to expand resource capacity to improve the processing power of the system.

\subsubsection{\textbf{Profiling and Comparison Template}}

Figure~\ref{fig:template-p3} illustrates the workflow that captures all 12 studies for \textit{Profiling and Comparison}~\citeS{sunardi2019mvc,fioravanti2016experimental,avritzer2018quantitative,haughian2016benchmarking, lung2014measuring,gonccalves2021monolith, alzboon2022performance, ngo2022evaluating, blinowski2022monolithic, avritzer2020scalability, hasselbring2020kieker}.
It contains four steps: 1) Alternative Architecture Acquisition, where the architectural solutions for an application are implemented or acquired; 2) Scenario Analysis, where typical workload scenarios for evaluating architectural solutions are identified and analyzed; 3) Dynamic Profiling, where run-time performance metrics in the identified workload scenarios are collected; and 4) Comparison, where the researchers make comparisons of the architectural solutions to provide empirical knowledge.

\textbf{Step 1: Alternative Architecture Solution Acquisition: } 

In seven of the studies, researchers implemented alternative architecture solutions for the same application~\citeS{fioravanti2016experimental, avritzer2018quantitative, lung2014measuring, gonccalves2021monolith, blinowski2022monolithic, avritzer2020scalability}. This approach provided control over the variables and facilitated the comparison of the impact of architectural changes on performance. For example, in~\citeS{fioravanti2016experimental}, a three-tier web application, \textit{E-Health Record}, was implemented using three different data storage systems. Similarly, the studies~\citeS{gonccalves2021monolith, blinowski2022monolithic, avritzer2020scalability} explored the same project's performance in  monolithic and micro-service architectures.

In four of the studies, researchers used off-the-shelf systems as their  subjects~\citeS{sunardi2019mvc, haughian2016benchmarking, alzboon2022performance, ngo2022evaluating}. In these cases, systems providing similar business functions but employing different architectures were selected for analysis. For instance, the study \citeS{haughian2016benchmarking} compared two different data storage architectures, i.e., \textit{Apache Cassandra} and \textit{MongoDB}. Extending this approach, \citeS{alzboon2022performance} contrasted two  cloud application architectures—--Ghost and WordPress. Furthermore, \citeS{ngo2022evaluating} carried out a performance evaluation of three  Function as a Service (FaaS) cloud platforms, including Amazon AWS Lambda, IBM, and Azure Cloud Function.

\textbf{Step 2: Scenario Analysis and Tuning:} This step tunes parameters that specify the input scenario or system resource, which impacts the performance of a system. This step is highly domain-specific. For studies that focus on web applications, \citeS{sunardi2019mvc} tuned the number of concurrent users; while~\citeS{lung2014measuring} controlled the number of requests~\citeS{ngo2022evaluating, lung2014measuring, gonccalves2021monolith, blinowski2022monolithic, avritzer2020scalability} and the queue size of messages~\citeS{zulkipli2013empirical}. Studies that focused on data storage systems controlled either the workload ~\citeS{haughian2016benchmarking, ngo2022evaluating} or the type of stored data~\citeS{fioravanti2016experimental}. More specifically, \citeS{haughian2016benchmarking} controlled two use case scenarios with the heavy-reading and heavy-writing. \citeS{fioravanti2016experimental} tuned the complexity of the data structures, e.g., tree-liked textual data or images, and different data replication. Tuning system resource, in terms of hardware configurations of the deployment environment, was another way to impact performance. This included 1) CPU allocation~\citeS{lung2014measuring}, 2) cache~\citeS{lung2014measuring}, 3) memory allocation~\citeS{lung2014measuring, avritzer2018quantitative}, and disk fraction~\citeS{avritzer2018quantitative}.

\textbf{Step 3: Performance Metrics Collection:} Researchers collect performance metrics using dynamic profiling tools and benchmarks~\citeS{sunardi2019mvc, avritzer2018quantitative, haughian2016benchmarking}.\textcolor{black}{Five studies~\citeS{gonccalves2021monolith, alzboon2022performance, ngo2022evaluating, blinowski2022monolithic}}, \citeS{sunardi2019mvc} used a \textit{Java} profiling tool, \textit{JMeter}, to collects the mean, maximal and minimal of response time and throughput. In \citeS{haughian2016benchmarking}, researchers used \textit{Yahoo Cloud Serving Benchmark (YCSB)} to monitor the throughput of \textit{Cassandra} and \textit{MongoDB}. Similarly, \textcolor{black}{\citeS{avritzer2020scalability} and} \citeS{avritzer2018quantitative} used an open source platform, named \textit{BenchFlow}, to automatically execute the performance tests and providing performance insights. \citeS{fioravanti2016experimental} collected the execution time based on the affiliated graphical user interface of the evaluated databases. \citeS{lung2014measuring} directly collected the performance metrics based on the built-in performance logs from \textit{Microsoft Windows XP Profession} operation system.

\textbf{Step 4: Comparison:} This step entails generating empirical knowledge via comparison. \textcolor{black}{In five out of eleven studies, specific architectural solutions consistently outperformed others~\citeS{fioravanti2016experimental, sunardi2019mvc, gonccalves2021monolith, alzboon2022performance}.} For instance, \citeS{fioravanti2016experimental} found that the \textit{NoSQL} systems---\textit{MongoDB} and \textit{Neo4j} --- delivered superior performance to the traditional \textit{MySQL + Hibernate} in both document-oriented and graph-oriented data contexts, with \textit{Neo4j} achieving a performance increase of 1.5 times on average, and \textit{MongoDB} accelerating by up to 33 times. \textcolor{black}{Likewise, studies such as \citeS{gonccalves2021monolith, avritzer2020scalability} compared two deployment architectures for web applications and consistently found the modern architecture to be superior to its older counterpart.} In \citeS{sunardi2019mvc}, the \textit{Slim Framework} outperformed the \textit{Laravel Framework} across all evaluated scenarios, although the authors conceded that the \textit{Laravel Framework} may be more suitable for large projects with numerous functions and external libraries. \textcolor{black}{Interestingly, \citeS{alzboon2022performance} found the WordPress blogging system to offer not only better performance than the Ghost blogging system, but also to be more stable.}

In the other 6 studies, different architectural solutions exhibited superior performance in specific scenarios~\citeS{haughian2016benchmarking, avritzer2018quantitative, lung2014measuring, blinowski2022monolithic, ngo2022evaluating}. 
\textcolor{black}{For instance, \citeS{haughian2016benchmarking} discovered that under a heavy-reading workload, \textit{MongoDB} marginally outperformed \textit{Cassandra}, whereas under a heavy-writing workload, \textit{Cassandra} was twice as efficient as \textit{MongoDB}. Similarly, \citeS{avritzer2018quantitative, ngo2022evaluating, lung2014measuring, blinowski2022monolithic} found that different architectural solutions achieved optimal performance under different deployment environments and workload scenarios.}

\subsubsection{\textbf{Study Template for Self-Adaption}} 

Figure~\ref{fig:template-p4} illustrates the general workflow of a self-adaptive framework: 1) Alternative Architectural Solution Acquisition, where the alternative architectural solutions are implemented or generated; 2) Performance Metrics Collections, where performance metrics are acquired by dynamic profiling or model-based performance prediction; 3) Decision Making, which determines whether and which alternative architectural solution should switch to; and 4) Execution, where the system enables the selected architectural solution.

\textbf{Step 1: Alternative Architectural Solution Acquisition:} In 4 studies~\citeS{lung2016improving, camara2020quantitative, gergin2014decentralized, peng2021parallel}, researchers manually implemented alternative architectural solutions for the same application. In the other 2 studies~\citeS{arcelli2020multi, ezzeddine2021design}, researchers firstly recovered the software models from an existing software system, and then proposed algorithms to re-configure the system model as alternative architectural solutions.

\textbf{Step 2: Performance Metrics Collection}: This step is to collect the performance metrics to support decision making in the next step. A studies~\citeS{arcelli2020multi} used model-based prediction methods to obtain the performance metrics. 
The other studies~\citeS{camara2020quantitative, lung2016improving, gergin2014decentralized, ezzeddine2021design, peng2021parallel} continuously monitored the performance metrics of a system. 


\textbf{Step 3: Decision Making:} This step is to analyze the performance metrics from the previous step to determine if an adaption is needed. 4 related studies' decision-making was based on switching policies~\citeS{lung2016improving, gergin2014decentralized, camara2020quantitative, peng2021parallel}. The switching policies are in three categories, including: 1) Threshold based policies, such as mean response time~\citeS{gergin2014decentralized, peng2021parallel} and data transmission rate~\citeS{camara2020quantitative}. Once the performance metric exceeds a threshold, the framework triggers a decision to adapt to suitable architecture solution. 2) Failure based policies~\citeS{lung2016improving}. In \citeS{lung2016improving}, the adaptation framework assumes that the system has a failure when the number of arrival requests stays normal or where the throughput drops to zero. And 3) Fluctuation and deviation based policies~\citeS{lung2016improving}. These policies compare the  monitored performance metrics with average metrics measured over a time period. Once it exceeds a specified percentage, the framework triggers a decision to adapt to an alternative architecture to better handle the requests.

The other two studies, \citeS{arcelli2020multi} and \citeS{ezzeddine2021design}, used a fitness algorithm to select the optimal solution from a set of candidate solutions, instead of relying on determined policies. 

\textbf{Step 4: Execution:} This step enables adaption to the most efficient architectural solution based on the decision from the previous step. In \citeS{lung2016improving}, the self-adaptive framework activates one of three alternatives---namely the \textit{DTC} architecture, the \textit{HS/HA} architecture, and the \textit{LF} architecture. Similarly, in \citeS{camara2013evolving}, the self-adaptive framework activates a more suitable data stream processing mechanism, based on the response time or the sampled delay rate. In \citeS{arcelli2020multi} and \citeS{camara2020quantitative}, the self-adaptive framework switches between alternative components deployment patterns of \textit{IoT} systems. 
In the other three studies~\citeS{gergin2014decentralized, ezzeddine2021design}, the self-adaptive framework re-configured  system connections~\citeS{ezzeddine2021design} or their resource allocation~\citeS{gergin2014decentralized}.
For example, in \citeS{gergin2014decentralized, peng2021parallel}, the self-adaptive framework manages the release of resources for system components to enable adaption.

Some steps like Model Acquisition, Architecture Acquisition or Implementation, and Performance Metrics Collection do appear commonly across templates. These commonalities underscore a shared foundation in the process of integrating architecture and performance analysis. As for practitioners, these commonalities serve as fundamental building blocks, enabling them to leverage existing knowledge and expertise across varied research projects. Moreover, the understanding of these shared steps can streamline the initiation of new projects, helping practitioners to focus on the unique aspects that align with their specific objectives.

\textbf{RQ3 Summary:} This RQ summarized the typical study templates of the four fundamental purposes from the 109 studies, which offer a guide for reproducing existing studies and advancing the state-of-the-art. Figures~\ref{fig:template-p1} to \ref{fig:template-p4} show the templates for \textit{Model-based Analysis}, \textit{Performance Anti-Pattern Analysis}, \textit{Profiling and Comparison}, and \textit{Self-Adaption}. Guidance regarding the key analysis components in the templates, including the architecture models, model annotation, performance models, and anti-pattern detection rules are provided.

\subsection{RQ-3: Available Tools}
\label{sec:rq4}
RQ3 identifies tools used in the studies to support the analysis of software architecture, performance, and their integration. Overall, in 13 studies, researchers indicated that the proposed approach or study process was fully automated with tool support~\citeS{li2016evaluating, eismann2018modeling, huber2016model, eismann2019integrating, gorsler2014performance, wu2020microrca, arcelli2020multi, eshraghian2015performance, walter2017expandable, du2015evolutionary, trubiani2015exploiting, trubiani2020visarch, arcelli2019exploiting}. In the other 12 studies, researchers explicitly mentioned that manual effort was required in the study process~\citeS{de2017model, trubiani2018exploiting, mazkatli2020incremental, trubiani2014guilt, sunardi2019mvc, nalepa2015model, cortellessa2014approach, gergin2014decentralized, wert2014automatic, yasaweerasinghelage2018predicting, avritzer2021multivariate, peng2021parallel}. In the remaining 61 studies, researchers did not elaborate on the tools for automation or the required manual effort.

We address RQ3 in two sub-RQs: 
\begin{itemize}
    \item \textit{RQ-3.1: What are the architecture and performance model analysis tools used in the literature?} Here we focus on any tools for constructing and analyzing the architecture and performance modules. We organize the tools' information based on their goals for model acquisition, model annotation, model transformation, and model analytical solution or simulation, which mostly align with the template for model-based prediction.  
    
    \item \textit{RQ-3.2: What are the profiling tools for collecting performance metrics, and what types of metrics are collected?} Here we focus on tools for collecting run-time performance metrics and mapping the tools to the different types of metrics being collected.
\end{itemize}

\subsubsection{RQ-3.1: What are the architecture and performance model analysis tools used in the literature?}

Table~\ref{tbl:tools} provides a summary of the tools used for the acquisition, annotation, transformation, and solution or simulation of the architecture and performance models. We would like to clarify that the table only enumerates the tools that we extracted from the 104 studies that are used in different analytical components of the study templates. However, we cannot assume the inter-changeability or interoperability among these tools. 

\textbf{1) Model Acquisition Tools:} Nine tools were mentioned in the literature for acquiring software architecture models. They are: \textit{UMLet}~\cite{li2016evaluating}, \textit{Enterprise Architect}~\citeS{li2016evaluating}, \textit{Palladio Bench}~\citeS{gorsler2014performance, trubiani2014guilt, yasaweerasinghelage2018predicting, happe2014stateful}, \textit{Eclipse Modeling Framework (EMF)}~\citeS{huber2016model, kross2017model}, \textit{Understand SciTool}~\citeS{zhao2020butterfly, zhao2020performance}. \textit{UMLet} and \textit{MagicDraw} allow users to manually draw UML diagrams. \textit{Enterprise Architect} not only supports manually drawing UML diagrams, but also reverse-engineering UML diagrams from system source code. \textit{Palladio Bench} is the default tool for reverse-engineering \textit{PCM} models from a code base. \textit{Enterprise Modeling Framework (EMF)} supports creating both \textit{PCM} models~\citeS{kross2017model} and \textit{DML} models~\citeS{huber2016model}. \textit{Understand SciTool} can recover the method-level call dependencies from a code base.

\textbf{2) Model Annotation Tools: } 
We identified three common tools for annotating the architecture model in preparation for performance prediction. They are the \textit{UML Profile for Schedulability, Performance, and Time (UML-SPT)}, \textit{UML Profile for Modeling and Analysis of Real-Time and Embedded systems (UML-MARTE)}~\citeS{gomez2018enabling, gomez2014performance, trubiani2014exploring, woodside2014transformation, arcelli2018performance, cortellessa2014approach, arcelli2019exploiting, requeno2017performance}, and \textit{Resource Demanding Service Effect Specification (RDSEFF)}~\citeS{brosig2014quantitative, brunnert2017continuous, noorshams2014enriching, yasaweerasinghelage2018predicting, kross2017model, trubiani2014guilt, happe2014stateful, mazkatli2020incremental}. Both \textit{UML-SPT} and \textit{UML-MARTE} can annotate UML diagrams. They can annotate performance information such as process duration, message size, branch action probability, network speed, and workload status. In addition, \textit{UML-MARTE} extends the annotation information to hardware aspects. For instance, \textit{UML-MARTE} can annotate a tag named ``HwDevice'', for distinguishing whether the hardware device is controlled by or interacts with a component in a software system. Similarly, the tag ``HwComponent'' annotates hardware components that physically contain other devices~\citeS{arcelli2019exploiting}. \textit{RDSEFF} is for annotating the \textit{PCM} models with information including request size (e.g. I/O resource demand), request type—read or write, the probability branch execution, and the file set that the request is operating on.

\textbf{3) Model Transformation Tools} We found 18 tools or techniques for transforming architecture models into either performance models or textual artifacts for \textit{model-based prediction}. Most tools work on transforming UML diagrams and PCM.

The UML models have nine transformation tools. Seven of them, \textit{XML Metadata Interchange (XMI)}~\citeS{geetha2011framework}, \textit{Query/View/Transformation (QVT) language}~\citeS{gomez2018enabling}, 
\textit{Janus Transformation Language (JTL)}~\citeS{arcelli2019exploiting}, \textit{PRIMA-UML}~\citeS{cortellessa2014approach}, and 
\textit{PUMA} \textit{PRIMA-UML}~\citeS{cortellessa2014approach, woodside2014transformation}  transform UML diagrams into \textit{Queuing Networks}.
Four tools are based on declarative transformation languages, which iteratively match the components in the source (i.e., architecture) model to the target (i.e., performance) model. \textit{SPE Editor (SPE·ED)}~\citeS{martens2011monolithic} not only supports the transformation from annotated UML diagrams into \textit{Queuing Networks}, but also support solving the generated \textit{Queuing Network} models with analytical techniques or with simulation~\citeS{smith2017spe}. Finally, \textit{ArgoSPE}~\citeS{gomez2014performance} is a tool to translate UML diagrams into \textit{Stochastic Petri Nets}.
There are five commonly used transformation tools for the \textit{PCM} models . \textit{PCM2LQN}~\citeS{brosig2014quantitative} and \textit{PCM2QPN}~\citeS{brosig2014quantitative} transform \textit{PCM} models to \textit{Layered Queueing Networks} and \textit{Queueing Petri Nets} respectively. \textit{KLAPER}~\citeS{gorsler2014performance} is used to transform \textit{PCM} into \textit{Queuing Petri Nets}. \textit{PCM2SimuCom}~\citeS{brosig2014quantitative, brunnert2017continuous, yasaweerasinghelage2018predicting, kross2017model, trubiani2014guilt} can automatically transform \textit{PCM} into simulation code. 

The remaining five tools are for transforming other architectural models. \textit{DML2SimQPN}~\citeS{huber2016model} and \textit{DML2LQNS}~\citeS{huber2016model} are for transforming \textit{DML} models into \textit{Layered Queuing Network (LQN)}, \textit{Queuing Petri Net (QPN)}, or simulation code. For transforming \textit{Æmilia}, a specific \textit{ADL}, \citeS{de2017model} transforms it to simulation code by using a declarative transformation language, \textit{Gra2Mol}, and to a performance model by using the tool, \textit{TwoTower}. For \textit{Calling Dependency Graph} model.

\textbf{4) Model Solution or Simulation Tools:} We identified 13 tools for solving a performance model or executing simulation code to obtain performance metrics for a system. 

There are five different tools for the analytical solution of \textit{Queuing Network} models. \textit{LQNS} is the default solver for a \textit{Layered Queuing Network (LQN)}; it is used in~\citeS{trubiani2020visarch, trubiani2014exploring, maddodi2020aggregate}. \textit{DLV Solver} and \textit{OPERA Solver} also solve  \textit{LQN} performance models. \textit{Java Modeling Tools (JMT)} is also a popular tool to approximate solutions for \textit{Queuing Network} models, used in~\citeS{arcelli2015performance, arcelli2018performance, trubiani2015exploiting, etxeberria2014performance}. Finally, \textit{SHARPE} is the default solver for \textit{Product Form Queuing Network}.

There are three tools for solving or simulating \textit{Petri Nets} models. \textit{QPN Solver} is for basic \textit{Queuing Petri Nets}~\citeS{eismann2019integrating}. \textit{GreatSPN} solves \textit{Generalized Stochastic Petri Net (GSPN)} models~\citeS{gomez2014performance, woodside2014transformation}. \textit{SimQPN} performs discrete-event simulation for \textit{Queuing Petri Net} models, used in~\citeS{eismann2018modeling, eismann2019integrating, huber2016model, walter2017expandable}.

The remaining five tools are for executing simulation code transformed from architecture models. \textit{SimuCom} is the most frequently used simulator for \textit{PCM} models~\citeS{brosig2014quantitative, brunnert2017continuous, yasaweerasinghelage2018predicting, happe2010parametric, kross2017model, trubiani2014guilt, happe2014stateful}. \textit{SimuLizar} is similar to \textit{SimuCom} but provides higher efficiency, scalability, and reliability. \textit{SimSo}~\citeS{nguyen2017parad} is for the simulation of the \textit{Architecture and Analysis Design Language (AADL)}. \textit{SMTQA} executes simulation code transformed from a \textit{UML Use Case Diagram}. \textit{SLAstic.SIM} executes simulation code transformed from \textit{Business Process Execution Language (BPEL)}.

\subsubsection{RQ-4.2: What are the profiling tools for collecting performance metrics, and what types of metrics are collected?}
All studies collected performance metrics. 33 out of 109 studies clearly mentioned the profiling tools that they used. 
In several studies, researchers did not use a dedicated profiling tool, but built their own profiling programs or collected performance metrics from operating system logs.

Six types of performance metrics were collected including 1) response time; 2) resource utilization; 3) throughput; 4) processing time; 5) latency; and 6) data transmission loss.
The three most commonly collected metrics are response time, resource utilization, and throughput. This finding is consistent with results from previous studies~\cite{aleti2012software, balsamo2004model}.

\begin{figure}[h]
  \centering
  \includegraphics[width=0.99\columnwidth]{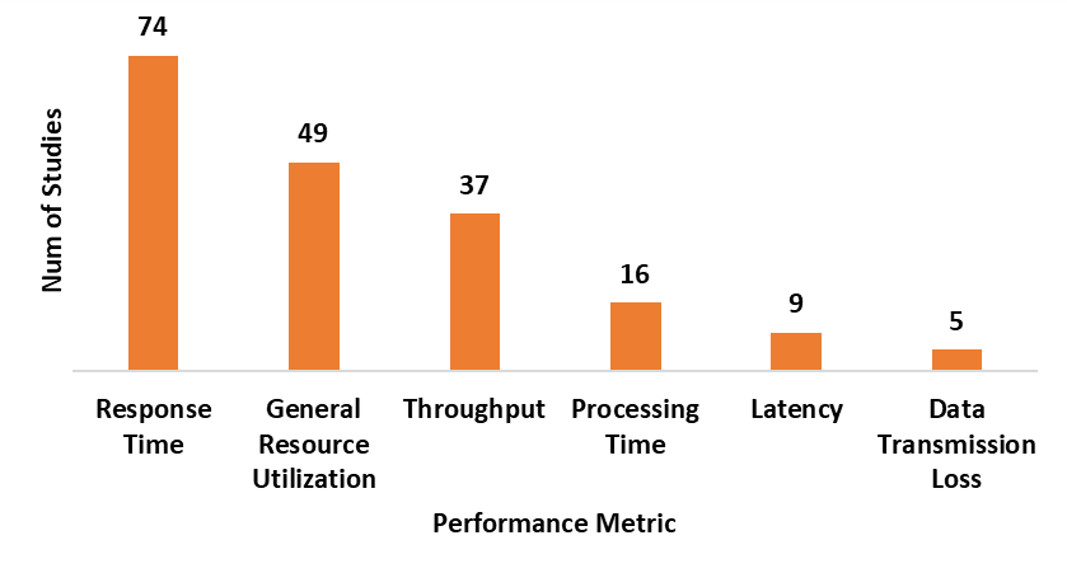}  
  \caption{Performance Metrics vs. Number of Studies}
  \label{fig:perf-metric-num}
\end{figure}

The array of studies reviewed in this paper employed an assortment of tools to gather metrics and oversee the performance of a broad variety of applications. For Java projects, profiling was often carried out using JMeter~\citeS{trubiani2018exploiting, arcelli2019exploiting, walter2017expandable, sunardi2019mvc, gonccalves2021monolith, alzboon2022performance, ngo2022evaluating, blinowski2022monolithic}, Java Flight Recorder~\cite{yasaweerasinghelage2018predicting}, and YourKit~\citeS{zhao2020butterfly, trubiani2018exploiting}. JMeter, being used in several studies, is capable of generating JUnit test cases automatically for more detailed profiling results. The Java Flight Recorder is notably light-weight, producing almost no performance overhead, whereas YourKit extends support for performance profiling to both Java and .NET projects. Java Management eXtension (JMX) has been employed for monitoring the performance of service-oriented Java applications.

For the performance monitoring of cloud-based applications with distributed clusters like Apache Hadoop, MRPerf~\citeS{wu2015exploring} and CloudWatch~\citeS{gergin2014decentralized} have proven to be useful. In ~\citeS{heger2013automated}, the tools ContiPerf and JUnitPerf were used to turn JUnit 4 tests into performance tests to measure performance within existing  tests, facilitating the smooth integration of performance testing into the unit testing process.


In terms of executing performance experiments and overseeing the performance of database systems like Cassandra and MongoDB, open-source frameworks such as Prometheus~\citeS{wu2020microrca}, BenchFlow~\citeS{avritzer2018quantitative, avritzer2020scalability}, and Yahoo Cloud Serving Benchmark (YCSB)~\citeS{haughian2016benchmarking} have been employed.
It's crucial to note that these tools have demonstrated versatility in their applicability, extending to various contexts and purposes beyond those described in this review.

    \begin{table*}[h]
	\centering
	\setlength{\tabcolsep}{11pt}
	\caption{Profiling Tools vs. Performance Metrics}{
		\resizebox{\linewidth}{!}{\begin{tabular}{|l|c|c|c|c|c|c|c|} 
				\hline
				Tool Name & Response Time & Resource Utilization & Throughput & Processing Time & Data Transmission Loss & Latency  \\  \hline
				\textit{JMeter} & \citeS{arcelli2019exploiting, trubiani2018exploiting, von2011performance, gonccalves2021monolith, alzboon2022performance, ngo2022evaluating, blinowski2022monolithic} & \citeS{trubiani2018exploiting, von2011performance} & \cite{sunardi2019mvc, trubiani2018exploiting} &  & \cite{sunardi2019mvc} & \citeS{trubiani2018exploiting} \\ 
				\textit{Java Flight Recorder} & \cite{yasaweerasinghelage2018predicting} & \cite{yasaweerasinghelage2018predicting} & & & & \\ 
				\textit{\textcolor{black}{YourKit}} & \citeS{trubiani2018exploiting}  & \citeS{trubiani2018exploiting}  & \citeS{trubiani2018exploiting} & \citeS{zhao2020butterfly} &  & \citeS{trubiani2018exploiting} \\ 
				\textcolor{black}{\textit{JMX}} & \citeS{litoiu2013performance} & \citeS{litoiu2013performance} & & & & \\ \hline
				\textit{MRPerf} & \citeS{wu2015exploring} & \citeS{wu2015exploring} & & & & \\ 
				\textit{\textcolor{black}{CloudWatch}} & \citeS{gergin2014decentralized} & \citeS{gergin2014decentralized} & \citeS{gergin2014decentralized} & & & \\ 
                \textit{ContiPerf} & \citeS{heger2013automated} & \citeS{heger2013automated} & & & & \\
                \textit{JUnitPerf} & \citeS{heger2013automated} & \citeS{heger2013automated} & & & & \\ \hline
				\textit{\textcolor{black}{WebGUI}} & \citeS{groenda2012improving} &  & & & & \\ 
				\textit{Nagios \& Cacti} & & \citeS{zulkipli2013empirical} & & \citeS{zulkipli2013empirical} & & \\ 
				\textit{\textcolor{black}{SNMP}} & \citeS{litoiu2013performance} & \citeS{litoiu2013performance} & & & & \\ \hline
				\textit{\textcolor{black}{Prometheus}} & \citeS{wu2020microrca} & \citeS{wu2020microrca} & & & & \citeS{wu2020microrca}  \\ 
				\textit{BenchFlow} & & \citeS{avritzer2018quantitative} & \citeS{avritzer2018quantitative} & & & \\ 
				\textit{YCSB} & & & \citeS{haughian2016benchmarking} & & & \\ \hline
	\end{tabular}}}
	\label{tbl:profiling-tools}
\end{table*}

\textbf{RQ4 Summary:} This RQ offers a catalog of tools for architecture analysis and for performance profiling curated from the 109 studies. This can help researchers and practitioners identify resources for their own research, or for reproduction studies. 
\section{Discussion of Future Directions}
\label{sec:implication}

\label{sec:future-directions}
\subsection{Lack of Reproducibility: } The lack of reproducible research has raised concerns in the software engineering community~\cite{boettiger2015introduction, madeyski2017would}. 
Although the literature that integrates software architecture and performance analysis has great potential, reproducibility of related studies remains a concern due to the lack of easy to reuse tools, benchmark datasets, and hands-on evaluations of these resources. Practical, hands-on evaluations of the tools and instruments could augment our study and provide more pragmatic insights into their effectiveness and usability. However, conducting such experiments involves many complexities, including the availability and usability of tools, and could constitute a series of separate studies in itself. This, we suggest, would be a valuable future direction for research in this field.

Therefore, increasing the reproducibility of related studies should be prioritized in future research. The community should emphasize the importance of including replication packages when publishing new studies.  This could help to increase the adoption of the proposed techniques. In addition, replication studies should be valued as an important type of scientific study in the community. This enables solid validation of new techniques by allowing study replication in multiple sites and by enforcing cross-project comparisons with the state-of-the-art.

\subsection{Challenges with Emerging Software System Domains:}  
Most studies focused on software systems in traditional domains as their evaluation subjects, such as E-commerce systems. Only four studies focused on cyber-physical system~\citeS{pinciroli2021model}, real-time embedded system~\citeS{gaudel2011ada, nguyen2017parad}, and Internet of Things system~\citeS{smith2017spe}, which posed challenges to the architecture and performance modeling resulted from the dynamism of the physical environment. 

New application domains such as block-chain systems, Web 3.0 systems, and virtual reality systems, may pose new performance challenges that are not sufficiently addressed in existing literature. For example, when building an architecture model of a blockchain system, one has to consider the consensus protocol that manages the transaction order, such as \textit{Proof of Work (PoW)} and \textit{Practical Byzantine Fault Tolerance (PBFT)}. In addition, new performance profiling techniques are needed to monitor performance metrics such as state updating time, consensus-cost time, and contract execution time~\cite{zheng2018detailed}, which cannot be collected using existing profiling instruments. As discussed in RQ4, most existing performance profiling instruments focus on measures of response time, throughput, and resource utilization, which may not meet the analysis needs of projects in specific domains. 

Therefore, another future direction is to develop new architecture modeling and performance analysis techniques that can address challenges in emerging software system domains.

\subsection{Limitations of the Evaluation Methodology:} 
A total of 30 studies only conducted a case study to show the feasibility of the proposed approach, without providing any specific evaluation criteria~\citeS{mazkatli2020incremental, martens2011monolithic,happe2014stateful,krogmann2010using,huber2016model,nalepa2015model,kwon2013mantis,brosig2011automated,li2016evaluating,brunnert2017continuous,noorshams2014enriching,hauck2010automatic,hauck2011ginpex,maddodi2020aggregate,eismann2018modeling,brosig2014quantitative,eismann2019integrating,walter2017expandable,happe2010parametric,requeno2017performance,werle2020data,yasaweerasinghelage2018predicting,brosig2014architecture,wu2015exploring,kross2017model,arcelli2017applying,arcelli2012antipattern,trubiani2014exploring,trubiani2018exploiting,arcelli2020multi}. 
And for studies that did focus on certain evaluation criteria, they mostly focused on technical effectiveness, such as accuracy of performance prediction, precision and recall of anti-pattern detection. 

In addition, there is a lack of involvement from industrial practitioners in the evaluations. As discussed in RQ5, only 29\% of the studies evaluated real-world systems~\citeS{wrzosk2012applying,gomez2014performance,gribaudo2018performance,requeno2017performance,wu2015exploring,kross2017model,eismann2019integrating,gomez2018enabling,woodside2014transformation,walter2017expandable,happe2010parametric,hill2010run,brosig2014architecture,camara2013evolving,arcelli2020multi,eismann2018modeling,koziolek2011peropteryx,haughian2016benchmarking,zhao2020butterfly,fioravanti2016experimental,chen2014detecting,zhao2020performance,kwon2013mantis,camara2020quantitative,fioravanti2017engineering,trubiani2018exploiting,maddodi2020aggregate}. 
In addition, these studies were mostly research-oriented---executing test cases of the system in lab environments to collect performance metrics. Evaluations that involve industrial practitioners can greatly benefit the research, promoting the adoption of the  research outcomes in practical settings.

\subsection{\textcolor{black}{Lack of ML/AI Techniques}:} 
Machine learning (ML) and artificial intelligence (AI) techniques have been increasingly adopted in software engineering activities~\cite{zhang2003machine, amershi2019software, washizaki2019studying}. However, in the literature we reviewed, only three studies~\citeS{camara2020quantitative, arcelli2020multi, kwon2013mantis} leveraged machine learning techniques. These studies applied machine learning for selecting influential performance factors and optimizing architectural solutions for performance self-adaptation.

Despite the rising prominence of ML/AI, its application in the area of software performance engineering is not sufficiently explored, due, in part, to the inherent complexities of these systems. They often lack the detailed information necessary for effective performance modeling, presenting unique challenges that traditional software architectures do not face.


In \textit{Model-based Prediction} studies, intricate architecture modeling, annotation, transformation, and solution processes often demand considerable manual effort. Researchers could explore leveraging ML and AI techniques to alleviate these challenges. In performance anti-pattern detection studies, detection rules are primarily heuristic, relying on practitioners' knowledge and expertise. The adoption of ML and AI techniques could enhance the accuracy of detection in practical settings. Similarly, \textit{Self-Adaption} studies, which largely depend on rules and thresholds for switching policies, could benefit from the integration of ML and AI techniques for more intelligent adaptation.

\subsection{Tentative Future Research Road-Map}

Based on the above discussion of limitations with existing studies, we would like to propose a tentative road map for researchers and practitioners to refer to and build upon:

\paragraph*{\textbf{Reproducibility}} Given the large quantity and high complexity of techniques and tools for integrating software architecture and performance analysis, reproducibility poses unique challenges to this direction. Here, we would like to propose the following research questions, derived from our research results:

\begin{enumerate}
    \item To what level the tools curated from RQ4 is actually available, supported, and usable for future research? It calls for a more in-depth investigation to actually retrieve, try, and test the tools as instructed in the original studies. 
    \item To what level the datasets used in the evaluation of the 104 studies are actually available and useful to the research community? As discussed in RQ5, most of the evaluation subjects are lab implementation and simulated systems. A small portion is real-world software projects. Examination of the availability and quality of the used dataset could serve for the construction of a shared benchmark, which is critical to reproducibility.
    \item To what extent the claimed results in existing studies can be reproduced, upon the confirmation of tool and data availability and quality? There is no guarantee that the same results could be achieved when repeated by a different research team. In particular, software architecture analysis is doomed subjective to an analyst's expertise; while dynamic performance profiling is volatile to the environment. These factors together add more challenges to reproducing prior results, underscoring its importance.
    \item To what extent that the various tools and techniques proposed in different studies are inter-operatable with each other? As presented in RQ3, different study templates share the same analysis components. As shown in RQ4, different concrete tools and techniques are employed for the same analytical components of the template in different studies. It remains largely open how the tools summarized in our RQ4 from different studies are actually interoperable with each other when plugged into the templates. This investigation would largely increase the flexibility and potential of architecture and performance analysis. 
     
\end{enumerate}

\paragraph*{\textbf{Emerging Domains}}

As presented in RQ1 (see Figure~\ref{fig:purpose}), researchers have attempted to provide customized solutions that meet the various demands of model-based analysis, resulting from specific architecture styles, specific system features, and certain analytical capabilities in special environments. However, with the emerging software applications domains encompassing all aspects of society, there is a significant gap between existing software architecture and performance analysis techniques and these emerging domains. Therefore, we would like to point to the following research questions for the community:

\begin{enumerate}
    \item How well do existing software architecture and performance analysis techniques apply to emerging domains, such as Web 3.0, Blockchain, Virtual Reality, etc.? This could be investigated based on the structure presented in RQ1 (see Figure~\ref{fig:purpose}). First, do the four fundamental purposes apply to emerging domains? Additionally, do the specific focuses of model-based performance analysis techniques cover the challenges of emerging domains?

    \item What are the unique challenges that are faced with the emerging domains? How existing techniques could be customized and enhanced to overcome them? We envision that emerging domains pose unique challenges that identify the limitations in existing studies. It would require multi-disciplinary research to picture their unique challenges for architecture and performance analysi.
\end{enumerate}

\paragraph*{\textbf{Practical Adoption}}

With the narrow focused evaluation subjects and factors, the practical adoption of the architecture and performance analysis techniques  remains questionable. Therefore, we would like to add the following research question to the future road-map:

What are the most significant challenges and impacting factors on the practical usage of architecture and performance analysis techniques? We assume that usability, learning curve, and cost-and-effectiveness all play critical roles in this realm. However, the most effective approach is to conduct human subject studies to reveal what challenges are actually encountered that will prevent the practical adoption. On top of that, engaging participants from practice to provide direct input through surveys, interviews, and focus groups would also be profoundly valuable.

\paragraph*{\textbf{AI Enabled Analysis}}

AI has gained widespread application in various domains, as well as in facilitating software engineering practice. Unique challenges arise in the application of AI for architecture and performance analysis. We would like to propose the following research questions that are customized to the unique challenges of this direction:

\begin{enumerate}
    \item How to shift architecture analysis' reliance on experts' expertise to AI? It is widely recognized that architecture analysis tends to be manual, labor-intensive, and error-prone. Therefore its success and effectiveness heavily rely on the experience and expertise of the analyst. It is yet an open question if AI would be able to provide a mitigation to this reliance, and to what extent.

    \item How to leverage AI for capturing the complexity and uncertainty in performance analysis? From our survey, we learned that performance analysis faces the unique challenges of uncertainty resulting from huge configuration space, composed of parameters and inter-play from the system as well as from the execution environment. Capturing the underlying complexity of such uncertainties often requires statistical analysis of the system's execution profile and environment. AI holds great promise in this area given that is is built upon powerful statistical analysis itself. 
\end{enumerate}

We acknowledge that the above road map is by no means comprehensive and authoritative. However, we believe that it provides a starting point for the community to build upon.
\section{Related Work}
\label{sec:related}

There are existing literature review studies that focus on software architecture and performance~\cite{aleti2012software, olabiyisi2010survey, becker2006performance, balsamo2004model, arcelli2020exploiting, becker2012model, falessi2010applying}. 
Here we discuss how our study differs from existing work.

Aleti \textit{et al.}~\cite{aleti2012software} presented a comprehensive systematic literature review on 188 studies (published between 1992 and 2011) that share the theme of software architecture optimization. The authors derived a taxonomy of architecture optimization concerns from these 188 studies, including: 1) performance (84 papers), 2) financial costs (74 papers), 3) reliability (71 papers), 4) availability (25 papers), 5) energy saving (18 papers), and 6) safety (4 papers). Similarly, the study by Falessi \textit{et al.}~\cite{falessi2010applying} summarized 11 empirical studies (between 2006 to 2008) that evaluate the methods, techniques, and tools that support architecture design, analysis, and review. This work focused on 1) the activities in software development (e.g., documentation design or maintenance evaluation), 2) research questions, and 3) study process of these 11 studies. 
Our work, compared with~\cite{aleti2012software} and ~\cite{falessi2010applying}, specifically focuses on the integration of architecture and performance analysis. Those papers have much broader scope, considering other quality concerns to motivate architecture optimization. In addition, our work focus on literature in the most recent decade, compared to the range of 1992 to 2011 in Aleti \textit{et al.}'s work~\citeS{aleti2012software} and the range of 2006 to 2008 in Falessi \textit{et al.}~\citeS{falessi2010applying}.

A number of secondary studies have already summarized the state-of-the-art in model-based performance prediction~\cite{olabiyisi2010survey, becker2006performance, balsamo2004model}. 
Balsamo \textit{et al.}~\cite{balsamo2004model} conducted a literature review that surveyed 15 studies in model-based performance prediction.  Their study focused on three aspects: 1) the identified software models and performance models; 2) the application in the software development life-cycle; and 3) the degree of automation. They found that \textit{UML Diagrams} are the most commonly used software architecture model; and \textit{Queuing Network} is the most preferred performance model. Most studies applied model-based prediction in the design phase. And, most of the approaches have a high degree of automation. The study \cite{olabiyisi2010survey} is a survey of performance models used for distributed software system architectures. It revealed that performance prediction approaches are mostly based on \textit{Queuing Networks}, \textit{Petri Nets}, \textit{Queuing Petri Nets}, and \textit{Hierarchical Performance Model}.
Similarly, \cite{becker2006performance} is a survey of 17 performance prediction studies, with a focus on component-based systems. This survey revealed that \textit{UML Diagrams} are the most commonly used software architecture model; and \textit{Queuing Network} is the most preferred performance model, which is consistent with our findings.  

Two surveys focused on self-adaptive systems. Arcelli presented a survey of 10 studies that apply performance modeling and assessment for supporting self-adaptive software systems~\cite{arcelli2020exploiting}. For each study, this survey analyzed the time of application (e.g., design time or running time), the architecture model (e.g., \textit{PCM}), the performance analysis methodology (e.g., analytical solution or simulated solution), as well as the applicability of the proposed tools. This survey pointed out that the state-of-art does not provide automated, performance-driven decision-making towards convenient alternative self-adaptive system architectures, by design-space exploration. This finding later motivated the author of the survey to propose a self-adaptive approach that is based on a genetic algorithm~\citeS{arcelli2020multi}, which is included in our literature. Becker \textit{et al.} presented another survey focusing on six self-adaptive approaches~\cite{becker2012model}. This study classified the six approaches based on their use scenarios as in design-time or run-time.
Our work differs from these studies~\cite{olabiyisi2010survey, becker2006performance, balsamo2004model,arcelli2020exploiting,becker2012model} in three ways: 1) Our  goal is to understand how software architecture and performance analysis can be integrated. Model-based prediction and self-adaptive systems are just two of the four study purposes that emerged from our literature set; 2) The   model-based prediction literature surveyed in our work is more recent and comprehensive. We reviewed the literature between 2010 and 2020, while prior work mostly focused on studies before 2010. And our work is also more comprehensive (with 94 studies); 3) For self-adaptive systems, existing surveys only find approaches that rely on a set of pre-defined policies to switch to an alternative architecture. However, we found two recent studies~\citeS{camara2020quantitative, arcelli2020multi} where researchers leverage machine learning techniques in making decisions for selecting suitable alternative architecture solutions. And, finally 4) We summarized the reference study templates of four research purposes, the available tools and instruments, and the evaluation methodology and domains for studies that integrate software architecture and performance analysis.


\section{Conclusion}
\label{sec:conclusion}

This paper presented a literature review of 109 studies that integrate software architecture and performance analysis. We addressed five research questions that revealed the study purposes and the facilitated development activities (RQ1), the reference research templates for different research purposes (RQ2), as well as the available tools and instruments (RQ3). The findings could provide reference and guidance for practitioners and researchers who are interested in software architecture, software performance, and the integration of the two. 

More specifically, in RQ1, we found that the majority of the surveyed studies focused on leveraging architecture modeling for performance analysis, while the rest addressed performance anti-pattern, performance profiling, and construction of self-adaptive architectures. We also discovered that the majority of the studies focused on addressing performance in the design phase. In RQ2, we summarized typical research templates of the four fundamental objectives of integrating software architecture and performance analysis, which could serve as quick guidance for researchers and practitioners to reproduce or advance existing studies. In RQ3, we summarized a catalog of the tools used in performance and architectural analysis, focusing on architecture modeling and performance profiling separately. This provides support for researchers and practitioners to identify tools that are relevant to their needs.

We also discussed potential future research opportunities based on limitations observed in existing studies. These include: 1) Lack of reproducibility---most studies lack replication package with available tools and dataset to reproduce their studies; 2) Challenges with emerging software domains---most studies focus on projects that are from traditional domains, such as e-commerce; while the focus on emerging domains such as Blockchain still needs more research; 3) Limitations with the evaluation methodology---most studies did not evaluate their techniques based on factors that impact the practical usage, such as usability; and 4) Lack of ML/AI techniques---modern machine learning and artificial intelligence techniques are not sufficiently used in the integration of architecture and performance modeling and analysis. Based on the identified limitations, we proposed a tentative future research road map for researchers and practitioners to refer to and build upon.



\bibliographystyle{unsrt}
\bibliography{refs}

\bibliographystyleS{unsrt}
\bibliographyS{myrefs}


\end{document}